\documentclass[twocolumn]{aastex62}
\graphicspath{{./}{figures/}}
\shorttitle{A Deep WISE Catalog}
\shortauthors{Schlafly et al.}
\usepackage{textcomp}
\usepackage{hyperref}
\usepackage{amsmath}
\usepackage{amsfonts}
\newcommand{\degree}{\ensuremath{^\circ}}
\newcommand{\mum}{\ensuremath{\mu}m}

\begin{document}

\title{The unWISE Catalog: Two Billion Infrared Sources from Five Years of WISE Imaging}

\author[0000-0002-3569-7421]{Edward F. Schlafly}
\affiliation{Lawrence Berkeley National Laboratory \\ 
One Cyclotron Road \\
Berkeley, CA 94720, USA}
\affiliation{Hubble Fellow}

\author[0000-0002-1125-7384]{Aaron M. Meisner}
\affiliation{National Optical Astronomy Observatory \\
950 N. Cherry Avenue \\
Tucson, AZ 85719, USA}
\affiliation{Lawrence Berkeley National Laboratory \\ 
One Cyclotron Road \\
Berkeley, CA 94720, USA}
\affiliation{Hubble Fellow}

\author[0000-0001-5417-2260]{Gregory M. Green}
\affiliation{Kavli Institute for Particle Astrophysics and Cosmology \\
Physics and Astrophysics Building, 452 Lomita Mall \\
Stanford, CA 94305, USA}
\affiliation{Porat Fellow}

\begin{abstract}
We present the unWISE Catalog, containing the positions and fluxes of roughly two billion objects observed by the Wide-field Infrared Survey Explorer (WISE) over the full sky.  The unWISE Catalog has two advantages over the existing WISE catalog (AllWISE): first, it is based on significantly deeper imaging, and second, it features improved modeling of crowded regions.  The deeper imaging used in the unWISE Catalog comes from the coaddition of all publicly available 3$-$5 micron WISE imaging, including that from the ongoing NEOWISE-Reactivation mission, thereby increasing the total exposure time by a factor of 5 relative to AllWISE.  At these depths, even at high Galactic latitudes many sources are blended with their neighbors; accordingly, the unWISE analysis simultaneously fits thousands of sources to obtain accurate photometry.  Our new catalog detects sources at $5\sigma$ roughly 0.7 magnitudes fainter than the AllWISE catalog and more accurately models millions of faint sources in the Galactic plane, enabling a wealth of Galactic and extragalactic science.  In particular, relative to AllWISE, unWISE doubles the number of galaxies detected between redshifts 0 and 1 and triples the number between redshifts 1 and 2, cataloging more than half a billion galaxies over the whole sky.
\end{abstract}


\section{Introduction} 
\label{sec:intro}

Expanding interest in cool objects and in high redshifts has driven continual progress in infrared astronomy, as enabled by tremendous improvements in detector sensitivity \citep{low2007}.  Within the past decade, the Wide-field Infrared Survey Explorer \citep[WISE;][]{wright2010} has provided an orders of magnitude leap forward relative to its predecessor IRAS \citep{wheelock1994}, mapping the entire sky at 3$-$22 microns with unprecedented sensitivity. As next-generation infrared missions like JWST, Euclid and WFIRST move forward, maximizing the value of existing infrared surveys like WISE will be critical.

By now, over 80\% of archival WISE data have been acquired via an ongoing asteroid-characterization mission called NEOWISE \citep{neowiser}. We are leading a wide-ranging effort to repurpose NEOWISE observations for astrophysics, starting by building deep full-sky coadds from tens of millions of 3.4 micron (W1) and 4.6 micron (W2)  exposures. Through our resulting ``unWISE'' line of data products, we have already created the deepest ever full-sky maps at 3$-$5 microns \citep{fulldepth_neo1, fulldepth_neo2, fulldepth_neo3}, generated a new class of time-domain WISE coadds \citep{tr_neo2, tr_neo3}, and performed forced photometry on these custom WISE coadds at the locations of more than a billion optically detected sources \citep{unwise_sdss_forcedphot, schlegel2015, dey2018}.

However, until now there has never been a WISE-selected catalog leveraging the enhanced depths achieved by incorporating NEOWISE data. Here we create and release such a catalog. Although WISE delivers exceptionally uniform and high quality imaging, its analysis requires careful application of appropriate computational techniques. At the WISE resolution of $\sim$6$\arcsec$, many sources substantially overlap others in the images, even at high Galactic latitudes where the fewest sources are detected.  This blending together of nearby sources renders many standard photometry codes unusable. The \texttt{crowdsource} crowded field point source photometry code \citep{Schlafly:2018}, recently developed for the DECam Galactic Plane Survey (DECaPS)\footnote{\url{http://decaps.skymaps.info}}, is well-suited to the task of modeling unWISE images, where nearly all objects are unresolved and blending is pervasive.

We have applied the \verb|crowdsource| photometry pipeline to deep unWISE coadds built from five years of publicly available WISE and NEOWISE imaging. The result is a catalog of $\sim$2 billion unique objects detected in the W1 and/or W2 channels, reaching depths $\sim$0.7 magnitudes fainter than those achieved by AllWISE \citep{Cutri:2013}. Our ``unWISE Catalog'' can therefore be considered a deeper W1/W2 successor to AllWISE with more than twice as many securely detected objects. This new catalog will have far-reaching implications, from discovering previously overlooked brown dwarfs in the solar neighborhood \citep[e.g.,][]{kirkpatrick2011} to revealing quasars in the epoch of reionization \citep[e.g.,][]{banados2018}.

In $\S$\ref{sec:wise} we recap the relevant history of the WISE and NEOWISE missions. In $\S$\ref{sec:unwise} we briefly highlight salient features of the unWISE coadds which form the basis of our unWISE Catalog. In $\S$\ref{sec:crowdsource} we describe our photometry pipeline. In $\S$\ref{sec:catalog} we provide an overview and evaluation of our resulting catalog. In $\S$\ref{sec:limitations} we discuss limitations of our current catalog processing and related avenues for future improvements. In $\S$\ref{sec:release} we describe the data release contents. We conclude in $\S$\ref{sec:conclusion}.

\section{WISE Overview}
\label{sec:wise}

Launched in late 2009, the WISE satellite resides in a $\sim$95 minute period low-Earth orbit. During the first half of 2010, WISE completed its primary mission by mapping the entire sky once in all four of its available channels, labeled W1 (3.4 microns), W2 (4.6 microns), W3 (12 microns) and W4 (22 microns), with a point spread function (PSF) of full-width at half-maximum (FWHM) of 6.1, 6.4, 6.5, and 12\arcsec. Over the following months, WISE ceased observations in W3 and W4 due to cryogen depletion, but nevertheless continued observing in W1 and W2 through early 2011 thanks to an asteroid-characterizing extension called NEOWISE \citep{neowise}. In 2011 February, WISE was placed into hibernation for nearly three years.  In late 2013, however, it was reactivated, and continued observations in W1/W2 observations as the NEOWISE-Reactivation mission \citep[NEOWISER;][]{neowiser}. The ongoing NEOWISE mission has now obtained nearly five full years (10 full-sky mappings) of W1 and W2 imaging, and has publicly released single-exposure images and catalogs corresponding to the first four of those years \citep[observations acquired between 2013 December and 2017 December;][]{neowise_supplement}.

Because NEOWISE is an asteroid characterization and discovery project, the mission itself does not publish any coadded data products of the sort that would maximize the raw NEOWISE data's value for Galactic and extragalactic astrophysics. AllWISE \citep{Cutri:2013} represents the most recent such set of coadded data products published by the WISE/NEOWISE teams, but was released at a time when only one fifth of the presently available W1/W2 data had been acquired. Because AllWISE already incorporates all available W3 and W4 imaging, we only construct catalogs in W1 and W2 in this work.

\section{unWISE Coadd Images}
\label{sec:unwise}

Ideally, our WISE cataloging would proceed by directly and jointly modeling all available W1/W2 exposures. However, doing so would be computationally intensive because these inputs represent $\sim$175 terabytes of data spread across $\sim$25 million single-exposure (``L1b'') images. As a computational convenience, our cataloging operates on a full-sky set of  36,480 coadded images totaling less than 1 terabyte in size. Specifically, we model deep unWISE coadds built from five years of single-exposure images in each of W1 and W2. These unWISE coadds uniformly incorporate all publicly available single-exposure images ever acquired in these bands, spanning the WISE and NEOWISE mission phases. The unWISE coaddition procedure is described in \cite{lang_unwise_coadds}, \cite{fulldepth_neo1, fulldepth_neo2} and \cite{fulldepth_neo3}. The five-year unWISE coadds used in this work are yet to be publicly released (Meisner et al. 2019, in prep.).

The unWISE coadds attain a 5$\times$ increase in total exposure time relative to the AllWISE coadds, so that we expect to achieve depths substantially beyond those attained by the AllWISE Source Catalog. Furthermore, the added redundancy of NEOWISE imaging allows our catalog to be relatively free of time-dependent systematics that were present in AllWISE, especially contamination from scattered moonlight.

The 36,480 unWISE coadd images are each $2048\times2048$ pixels in size with a pixel scale of 2.75\arcsec, covering about 2.5 square degrees.  Each image is identified by one of 18,240 \verb|coadd_id| values giving the location of the coadd on the sky, and a band (W1 or W2). The unWISE tile centers and footprints match those of the AllWISE Atlas stacks.

\section{Crowded Field Photometry Pipeline}
\label{sec:crowdsource}

Cursory inspection of even the highest-latitude, least crowded images reveals that many sources in the unWISE coadds are significantly blended with their neighbors.  Accordingly, fully taking advantage of the unWISE coadds requires a crowded-field photometry pipeline that simultaneously models the many blended sources in each field.

The WISE images are in many ways ideal for crowded-field analysis techniques.  A substantial challenge in modeling crowded fields is accurate determination of the PSF and its wings.  Fortunately, the WISE satellite has a very stable PSF owing to its location above the atmosphere.  Accordingly, we can adopt a nearly constant model for the shape of the PSF, and tweak only the very central region as necessary to improve the fit.

The WISE images are also ideal for crowded-field analysis techniques because of their relatively large $\sim 6\arcsec$ PSF FWHM.  This large PSF means that most distant galaxies are not resolved and can be adequately modeled as point sources, introducing only a small bias.  The unWISE Catalog analysis simply assumes that \emph{all} sources are point sources.  This tremendously simplifies the modeling relative to typical optical extragalactic surveys with $\sim 1\arcsec$ PSFs, where detailed modeling of the shapes of galaxies is required to match the observed images.

We use the \texttt{crowdsource} analysis pipeline to model the unWISE images \citep{Schlafly:2018}.  This pipeline simultaneously models all of the sources in each $512\times512$ pixel region of an unWISE tile, optimizing the positions and fluxes of the sources as well as the background sky to minimize the difference between the observed image and the model.  This pipeline was designed for ground-based optical images, but application to WISE images poses few additional problems.

\begin{figure}[htb]
\begin{center}\includegraphics[width=\columnwidth]{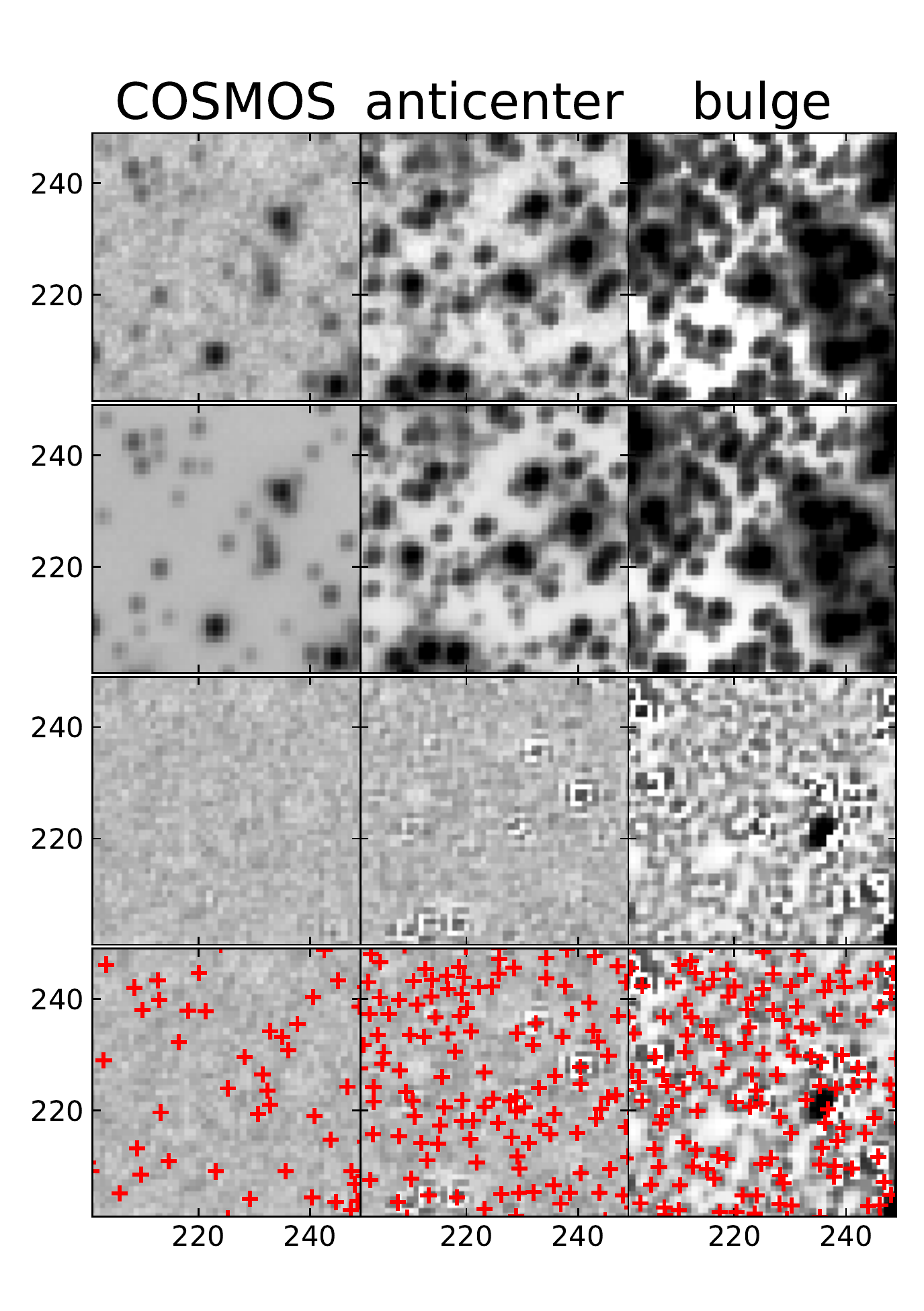}\end{center}
\caption{
\label{fig:crowdsourceexample}
\texttt{crowdsource} modeling results in three fields with very different source densities: the high-latitude COSMOS field (left), the Galactic anticenter (middle), and the Galactic bulge (right).  From top to bottom, the rows show the unWISE coadded images, the \texttt{crowdsource} model, the residuals, and the residuals with the locations of cataloged sources overplotted.  Except for in the bulge field, shot noise accounts for a substantial fraction of the residuals.  On the other hand, in the densest regions, like the bulge, the residuals are completely dominated by unresolved sources and challenges in sky subtraction.  The bulge field is stretched $10\times$ less hard than the other two fields.  In all cases, the model images account for most of the flux in the real images, though clearly in the bulge case significant residuals remain.  All images are W1.
}
\end{figure}

Figure~\ref{fig:crowdsourceexample} shows examples of the \texttt{crowdsource} modeling in three fields of very different source densities.  The first column shows a portion of the COSMOS field, at high Galactic latitude; the second shows a portion of the Galactic anticenter; and the third shows the Galactic bulge.  At high latitudes, and even directly in the Galactic plane, the \texttt{crowdsource} model (second row) is an excellent description of the unWISE images (first row).  The differences between the two (third row) clearly come substantially from the shot noise in the images, at least outside the cores of bright stars.  Often inspection of the residual image at the locations of the detected sources (fourth row) shows no coherent residual signatures.  

Meanwhile, in the Galactic bulge (third column), the story is very different.  The large WISE PSF coupled with the tremendous number of sources and insensitivity to dust extinction make this field very confused.  While the bulge image (top row, right) and model (second row, right) are qualitatively in good agreement, the residuals (third row, right) are entirely dominated by coherent structures stemming from the incomplete identification of significant sources in the field.  It may be possible to do better here by allowing \texttt{crowdsource} to more aggressively identify unmodeled stars overlapping brighter neighbors, but even very slight errors in the PSF model will have substantial effects on the residuals in fields as dense as this one, rendering the results of a more aggressive source identification uncertain at best.

A modest number of modifications and improvements were made to the \texttt{crowdsource} pipeline to allow it to model WISE images.  These changes included changes to the PSF modeling, the mosaicing strategy, the sky modeling, and the treatment of nebulosity and large galaxies.  Additionally, a new diagnostic field \texttt{spread\_model} was added to the pipeline outputs, which can help determine the size of detected objects.

\subsection{PSF modeling}
\label{subsec:psf}

The most significant addition to \texttt{crowdsource} was a PSF modeling module specifically designed for WISE.  The unWISE PSF module\footnote{\url{https://github.com/legacysurvey/unwise_psf}} provides a $325\times325$ pixel PSF, developed using bright isolated stars \citep{wise_dust_map}.  This PSF extends far into the wings of the WISE PSF, which has a full-width at half-maximum of only about 2.5 pixels.  It includes details of the PSF like diffraction spikes, optical ghosts, and the optical halo.  This model was originally designed for application to the unWISE single-exposure images.  The unWISE coadd images sum many single-exposure images together, necessitating changes to the original unWISE PSF model.  Near the ecliptic poles, the single-exposure images contributing to a given coadd image span a wide range of different detector orientations relative to the sky, leading the final coadd image PSF to be blurred over a range of azimuth.  This is modeled by transformation of the PSF to polar coordinates and convolution with a boxcar kernel in azimuth, with the width and position of the boxcar kernel dependent on ecliptic latitude and longitude.  The PSF is then projected back to cartesian coordinates.

The resulting model gives an excellent description of the observed PSF at any particular point in the survey.  The convolution process is somewhat expensive, however.  To save time, a grid of these PSF models is generated for each unWISE tile.  At low ecliptic latitudes, the coadd PSF is essentially constant over each unWISE tile, and a relatively coarse grid is adequate.  At high ecliptic latitudes (within about 20\degree\ of the ecliptic poles), the mean spacecraft orientation and the width of the range of orientations vary more significantly across each coadd tile.  This requires a denser grid of PSFs to model.  Within 5\degree\ of the north ecliptic pole, one PSF per $128\times128$ pixel region is generated.  The \texttt{crowdsource} pipeline linearly interpolates between these PSFs when generating a model for a source on a tile.

These PSFs are reasonably accurate within a hundred pixels of the PSF center.  Beyond this region, subtleties with the world-coordinate system of the coadds versus the PSF become important.  Moreover, some coherent residuals in the PSF cores are also apparent.  The \texttt{crowdsource} PSF module addresses these by, at each \texttt{crowdsource} modeling iteration, finding the median residual over all bright, unsaturated stars in the inner $9\times9$ pixel PSF core, and adding it to the PSF model.  Typical remaining residuals stem from imperfect subpixel interpolation of the PSF, as indicated by ringing in the residuals around bright stars, as seen in the anticenter residual image in Figure~\ref{fig:crowdsourceexample}.  Further effort could eliminate these artifacts; the huge number of stars in each image clearly provide more than enough information to constrain the subpixel shape of the PSF.

\subsection{Mosaicing Strategy}
Modeling crowded images can require substantial amounts of computer memory.  Individual unWISE coadd images in the inner Galaxy contain more than 200,000 detected sources.  \texttt{crowdsource} constructs a sparse matrix containing the PSF for each of these sources.  Bright sources can require PSFs extending out to $299\times299$ pixels, for $\sim$90,000 values per source, so a naive approach could require $\sim200$ GB of memory for the sparse matrix alone in this extreme case.

To make this more manageable, \texttt{crowdsource} splits each unWISE tile into $512\times512$ pixel subimages, with an additional 150 pixel border on each side.  In the original incarnation of \texttt{crowdsource} for the DECam Plane Survey, these final catalogs for each subimage were created in turn.  In order to prevent duplicate source detection in overlap regions, the analysis of later subimages included fixed sources at the locations of sources detected in earlier subimages.

In unWISE, however, source detection occurs for the entire coadd simultaneously.  The parameters of these sources are then optimized on each image subregion.  Finally, again over the entire coadd, the PSF is refined and source detection is repeated.  This new approach preserves the computer-memory advantages of the former approach while allowing PSF modeling to be performed on the entire coadd and more gracefully handling sources in the overlap regions between subimages.

\subsection{Sky Modeling}
A slightly different approach for sky modeling was taken for WISE than for the DECam Plane Survey.  In the DECam Plane Survey, the sparse linear algebra solver was allowed to adjust the overall sky level simultaneously with the fluxes of the stars.  In dense regions, this allows an an initial sky overestimate (due to the presence sources in the image) to be improved by simultaneously decreasing the sky and increasing the fluxes of the sources.

This global approach has the disadvantage that isolated, very poorly fitting regions of the image can significantly drive the sky estimate over the entire image.  The WISE coadds feature a larger dynamic range than DECam images, making it easier for small residuals in the cores of bright stars to make outsize contributions to the likelihood in the fitting.  This problem was addressed by eliminating the overall sky parameter from the sparse linear algebra fit.  The sky, however, is still improved at each iteration by median filtering the residual image, as in \citep{Schlafly:2018}.  This area of the pipeline offers room for improvement; see \textsection\ref{subsec:skysubtractionlimitation}.

\subsection{Nebulosity and large galaxies}
\label{subsec:nebulosity}
Reflected starlight and thermally emitted light from dust grains can add a diffuse component with rich small-scale structure to observed infrared images.  The \texttt{crowdsource} modeling assumes that all flux not explained by a smooth sky model must be attributable to point sources.  Similarly, large galaxies present in the WISE imaging will be split into many point sources unless preventative measures are taken.

To address both of these cases, we identify nebulosity and large galaxies ahead of time and mark these regions.  During source detection, candidate sources found in regions marked as containing nebulosity are then required to be sharp; significantly blended objects in regions with nebulosity will not be modeled.  In \citet{Schlafly:2018}, these sources were additionally required to not overlap any neighbors substantially, but for WISE we removed this constraint.  In the optical, regions with nebulosity are usually sparsely populated with stars; the extinction associated with the gas and dust limits the depth of the survey in nebulous areas.  In contrast, the infrared light observed by WISE is hardly extinguished by dust, causing application of a strong blending criterion to source detection to eliminate many real sources.

Sources found in regions containing large galaxies, on the other hand, are not required to be sharp, but \emph{are} required to not overlap any neighboring source substantially.  This is appropriate since peaks corresponding to large galaxies will naturally be extended and fail a sharpness cut, but only a single source should be associated with these galaxies.  Requiring new sources to not overlap existing ones significantly discourages \texttt{crowdsource} from splitting these galaxies into many point-source components.  The delivered fluxes, however, will still be inaccurate, as these large galaxies are modeled as single point sources.

Regions of nebulosity are identified with the same machine learning approach as in \citet{Schlafly:2018}.  A few minor changes were made, however.  First, the convolutional neural network was trained with $256\times256$ pixel images, instead of the $512\times512$ images used in \citet{Schlafly:2018}.  Second, a slightly shallower convolutional neural network (see Appendix \ref{app:neural-network-structure}) was adopted.  Finally, the neural network was retrained using a set of WISE images rather than optical images.  The neural network nebulosity classifications are available as part of the mask images in the Data Release (\textsection\ref{sec:release}).

The HyperLeda catalog \citep{Makarov:2014} of large galaxies was used to mark regions containing galaxies resolved in WISE ($D_{25} > 10\arcsec$).  These regions are marked in the unWISE mask images; see Meisner et al. (2019, in prep.) for details of their implementation in unWISE.  HyperLeda is also used to select large galaxies for further analysis in the Legacy Survey Large Galaxy Atlas\footnote{\url{https://github.com/mostakas/LSLGA}}.

\subsection{\texttt{spread\_model}}
\label{subsec:spreadmodel}
The unWISE catalog modeling assumes that all sources are point sources; their shapes are completely described by the modeled PSF.  Nevertheless, roughly half of all sources in the catalog are galaxies, which are not point sources.  The size of a source can be a useful diagnostic for identifying galaxies or problems with the modeling; for example, a single catalog source that corresponds to two overlapping point sources should be slightly larger than the PSF.  Additionally, instrumental effects, like trends in PSF shape with magnitude (for example, as caused by the brighter-fatter effect, \citealt{Downing:2006, Antilogus:2014}), or with color (e.g., PSF chromaticity, \citealt{Cypriano:2010}) can be identified through measurements of source size.  The \texttt{crowdsource} pipeline now follows the lead of \texttt{SExtractor}, computing \texttt{spread\_model} as a measure of a source's size \citep{Bertin:1996, Desai:2012}\footnote{The \texttt{crowdsource} \texttt{spread\_model} is corrected from the formulation in \citet{Desai:2012} following the \texttt{SExtractor} documentation so that in the absence of noise point sources have \texttt{spread\_model} equal to 0.}.  In combination with its uncertainty \texttt{dspread\_model}, this term roughly indicates how significantly increasing the size of the PSF would improve the fit.

\section{unWISE Catalog}
\label{sec:catalog}

\begin{figure*}[htb]
\begin{center}\includegraphics[width=\textwidth]{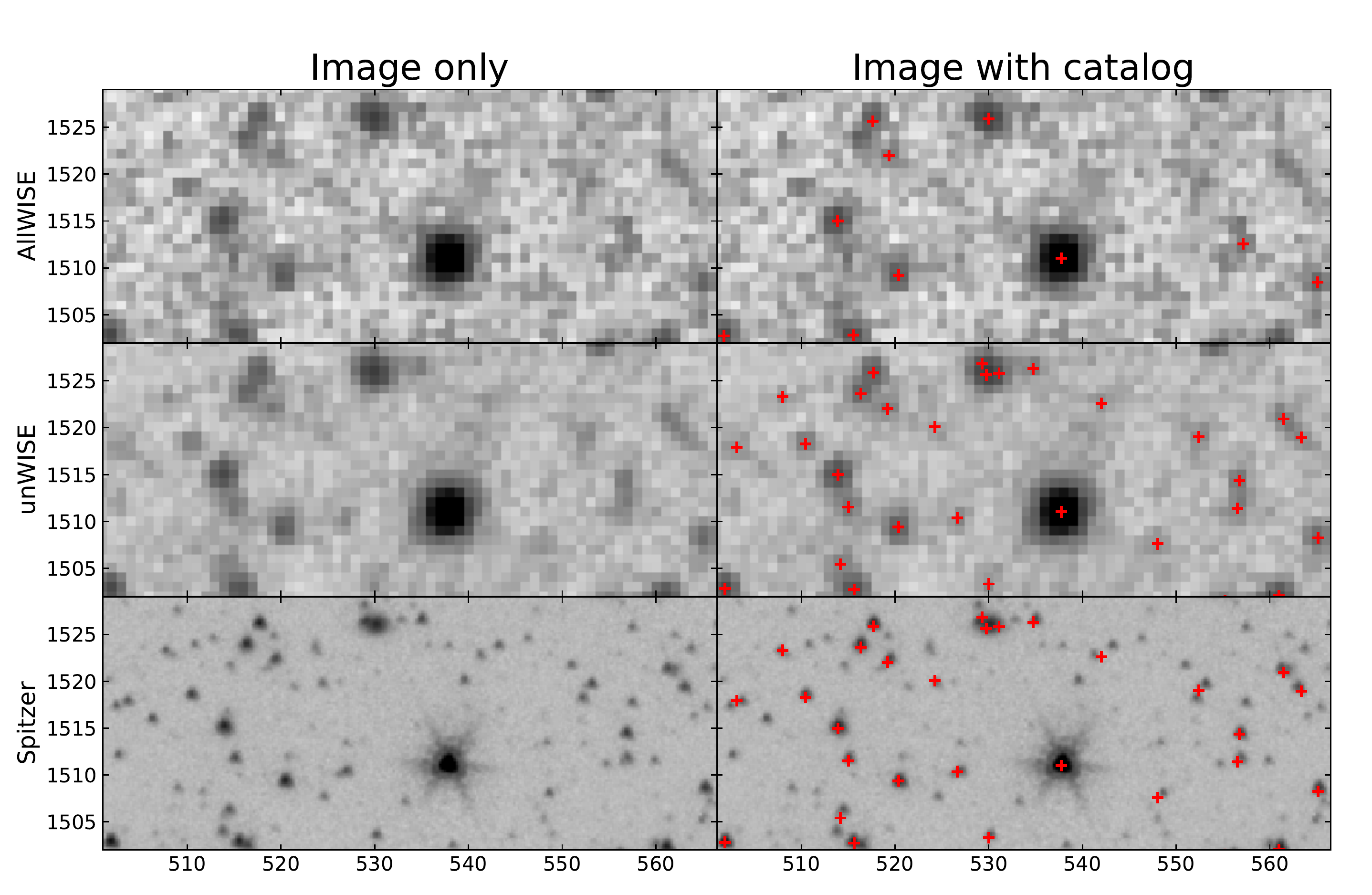}\end{center}
\caption{
\label{fig:auscompare}
AllWISE and unWISE compared with much deeper, higher-resolution imaging from Spitzer-COSMOS, for a small portion of the COSMOS field.  The three rows show 3.4\mum\ imaging of the same small patch of high-latitude sky from AllWISE (top), unWISE (middle), and Spitzer-COSMOS (right, 3.6\mum).  The left column shows only the images, while the right column overplots the $5\sigma$ catalog entries from AllWISE (top) and unWISE (middle, bottom).  The deeper unWISE stacks clearly allow many more sources to be detected, and the \texttt{crowdsource} catalog well describes these.  Nevertheless, comparison with the Spitzer imaging reveals clear examples of unidentified sources (for instance, near (503, 1518)) and resolved sources that are split into multiple point sources (for instance, near (530, 1526)).  Axis units are WISE pixels on \texttt{coadd\_id} \texttt{1497p015}.
}
\end{figure*}

The unWISE catalog uses the deep unWISE coadds to detect sources a factor of two fainter than possible in AllWISE, detecting a total of about 2 billion objects.  It couples this with the \texttt{crowdsource} source detection and characterization pipeline to model significantly blended sources.  Figure~\ref{fig:auscompare} illustrates the improvement realized by this combination, showing the AllWISE and unWISE images with the corresponding catalogs, and finally the much deeper, higher-resolution S-COSMOS imaging from Spitzer \citep{Sanders:2007}.

The dramatic improvement in depth realized by the four additional years of NEOWISE-Reactivation imaging is immediately apparent comparing the upper left and middle left panels of Figure~\ref{fig:auscompare}.  Additionally, the importance of modeling blended sources is clear.  Many sources within the unWISE coadd significantly overlap one another, even in this high-latitude field.  Comparison of the unWISE coadd image with the corresponding catalog (middle right) suggests that \texttt{crowdsource} has done a good job identifying sources in the field.  The much deeper Spitzer imaging (bottom row) generally confirms this impression, though there are clear instances of cases where \texttt{crowdsource} has split galaxies into multiple sources (for example, the large galaxy near (530, 1526)), as well as cases where it has failed to split overlapping, unresolved, faint sources into their components (for example, the pair of sources near (503, 1518)).  The former case is not particularly troubling; there are not that many galaxies resolved by WISE, and the ones that exist are usually securely identified in higher-resolution imaging.

\begin{figure*}[htb]
\begin{center}\includegraphics[width=\textwidth]{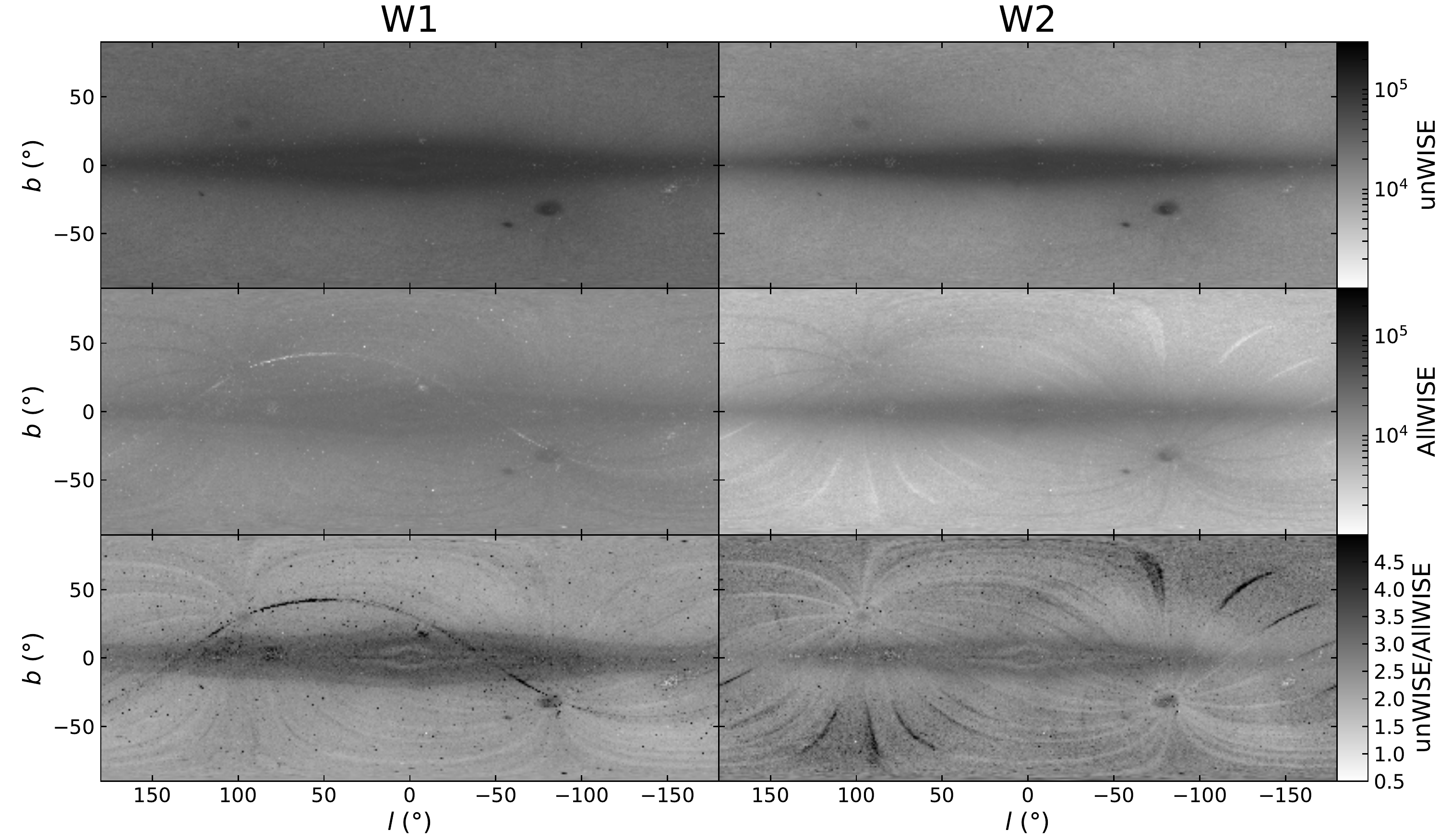}\end{center}
\caption{
\label{fig:numdenscompare}
Number density of sources per square degree cataloged by unWISE and AllWISE.  The top row shows the unWISE catalog densities; the middle row shows AllWISE; and the bottom row shows the ratio of the two.  unWISE detects more than $2\times$ as many sources as AllWISE at high latitudes, and more than $3\times$ as many sources as AllWISE at low latitudes.  Scattered light from the Moon leads to significant variations in the AllWISE number densities in W2 at high latitudes; these are mostly eliminated in unWISE owing to the greater number of observations, which allows all parts of the sky to be observed in conditions free from significant scattered moonlight.  Isolated regions of elevated unWISE/AllWISE source count ratio usually correspond to the $\sim2000$ brightest stars in the sky, where unWISE tends to detect too many sources (incorrectly identifying features in the wings of the PSF of very bright stars as separate sources) and AllWISE tends to detect too few sources (missing many objects in the wings of bright stars).  Regions with significant nebulosity show a deficit of objects in both unWISE and AllWISE---for instance, in Orion near $(-150\degree, -20\degree)$.
}
\end{figure*}

Unsurprisingly, the unWISE catalog contains many more objects than the AllWISE catalog.  Figure~\ref{fig:numdenscompare} shows the number densities of $5\sigma$ sources per square degree detected in unWISE (top) as compared with AllWISE (middle), and the ratio of the two number density maps (bottom).  In typical high-latitude locations, unWISE includes about $2\times$ as many objects in W1 and $2.5\times$ as many objects in W2.  In the Galactic plane, the difference is more dramatic, with unWISE cataloging 3--3.5$\times$ more sources than AllWISE, owing to the aggressive identification of blended stars in \texttt{crowdsource}.

The ratio maps additionally show some isolated regions of elevated unWISE/AllWISE detections at high latitudes.  These tend to surround the brightest $\sim2000$ stars in the sky.  The unWISE Catalog detects too many sources in these regions, incorrectly interpreting errors in the wings of the PSF model as faint sources.  AllWISE, on the other hand, detects too few sources in these regions, because it overestimates the sky in the vicinity of bright objects and because it relegates sources too near bright objects to the ``reject'' table.

Regions of substantial nebulosity also are evident in Figure~\ref{fig:numdenscompare} as having low source densities in unWISE and AllWISE; the clearest example is in the vicinity of the Orion Molecular Cloud near $(l, b) = (-150\degree, -20\degree)$.  The unWISE Catalog imposes an additional criterion on candidate sources in these regions to limit the number of spurious objects detected (see \textsection\ref{subsec:nebulosity}).  However, significant numbers of real sources are also excluded by this criterion.

Clear striping along lines of constant ecliptic longitude are especially evident in the number count ratios (Figure~\ref{fig:numdenscompare}, bottom).  These stripes usually correspond to periods of the survey where the Moon was near the part of the sky being observed, scattering light into the images and reducing their depth.  These features are absent from the unWISE catalog because the additional years of NEOWISE observations provide moon-free imaging of these regions.  The particularly prominent stripe in the W1 ratio map lies at ecliptic longitude $\lambda \approx 240\degree$, which was observed at the beginning of the 3-band Cryo phase.  Sources in AllWISE in this region may be reported as having zero uncertainty, leading them to be excluded from Figure~\ref{fig:numdenscompare}; see \textsection II.2.c.ii of \citet{Cutri:2013} for details.

\subsection{Completeness and Reliability}
\label{subsec:completeness}
To assess the completeness and reliability of our catalog in a representative sky location, we compare against Spitzer data in the COSMOS region.  COSMOS is at high Galactic latitude, and so serves as a typical extragalactic field.  Moreover, it is at low ecliptic latitude, so the WISE imaging is typical in depth.  For our purposes, S-COSMOS is the preferred Spitzer data set in the COSMOS region, as it is much deeper than our unWISE catalog and covers the entire 2 square degree COSMOS footprint. Our analyses compare WISE W1 and W2 against Spitzer ch1 and ch2, respectively, since these pairs of bandpasses are quite similar, although not identical. In both our completeness and reliability analyses, we use a Spitzer-WISE cross-match radius of 2$''$. Portions of four unWISE \verb|coadd_id| footprints contribute to the analysis: \verb|1497p015|, \verb|1497p030|, \verb|1512p015|, \verb|1512p030|. All completeness and reliability values presented are differential.

\begin{figure*}
\begin{center}\includegraphics[width=\textwidth]{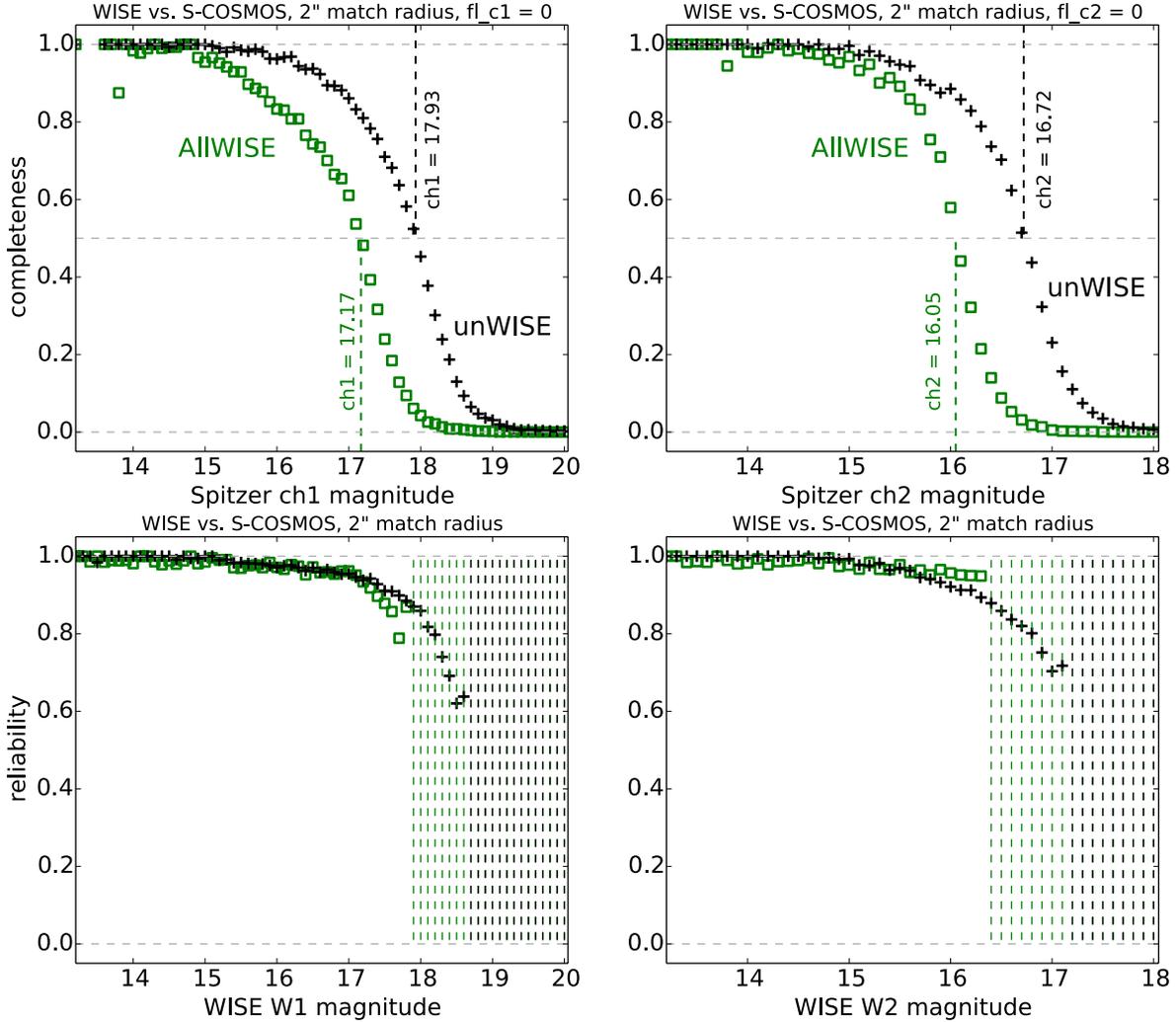}\end{center}
\caption{
\label{fig:completeness_reliability}
Summary of our completeness and reliability analysis based on a comparison against the Spitzer S-COSMOS data set over $\sim$2 square degrees of extragalactic, low ecliptic latitude sky. Top: differential completeness as a function of Spitzer magnitude for AllWISE (green squares) and the unWISE Catalog (black plus marks). The unWISE Catalog hits 50\% completeness $\sim$0.7 magnitudes fainter than AllWISE in both bands. Bottom: differential reliability as a function of WISE magnitude, with the same marker symbols/colors as above. Vertical dashed green lines indicate magnitude bins with $< 10$ AllWISE \texttt{w?snr} $\ge 5$ sources in the relevant band. Vertical dashed black lines indicate magnitude bins containing $< 10$ sources with signal-to-noise $\ge 5$ in either AllWISE or unWISE. Note that although the numerical values along the horizontal axes of the upper and lower panels are aligned, their units are different (Spitzer magnitudes in the upper panels and WISE magnitudes in the lower panels).}
\end{figure*}

To measure the unWISE catalog's completeness, we wish to compare against a highly reliable Spitzer catalog. For this reason, we take as a sample of ``true'' Spitzer sources the subset of S-COSMOS catalog entries with \verb|fl_c?| = 0 and \verb|flux_c?_4| $>$ 0 in the relevant channel. One consequence of our decision to cut on the S-COSMOS quality flag \verb|fl_c?| is that sources brighter than $\sim$13 mag in ch1 and ch2 are rejected, preventing our completeness analysis from reaching bright magnitudes. We also remove Spitzer sources with locations marked by \verb|flags_info| bit 1 as being associated with a WISE-resolved galaxy -- this effectively restricts our analysis to pointlike sources. The top two panels of Figure \ref{fig:completeness_reliability} show the completeness as a function of Spitzer  magnitude, for both our unWISE Catalog and the AllWISE Source Catalog. Because AllWISE performs forced photometry in every WISE band for sources detected in any band, we have restricted to sources with signal-to-noise ratio greater than 5 in the corresponding band.  This provides a fair comparison against the unWISE Catalog, which does not perform forced photometry. In W1, we find that the unWISE Catalog is 0.76 mags deeper than AllWISE, with the former (latter) reaching 50\% completeness at ch1 = 17.93 (17.17) mag. In W2, we find that the unWISE Catalog is 0.67 mags deeper than AllWISE, with the former (latter) reaching 50\% completeness at ch2 = 16.72 (16.05) mag. In AB, these unWISE catalog depths are 20.72 (19.97) mag in W1 (W2).


Given that our unWISE catalog benefits from $\sim$5$\times$ enhanced integer coverage relative to AllWISE, one might naively expect that we should reach 2.5log$_{10}\sqrt{5}$ = 0.87 mags deeper in both W1 and W2, assuming that the pre- and post- hibernation data are of identical quality. In detail, there are two main factors that result in a lesser depth enhancement. First, the WISE PSF is a few percent broader in both W1 and W2 in post-hibernation imaging relative to pre-hibernation imaging\footnote{\url{http://wise2.ipac.caltech.edu/docs/release/neowise/expsup/sec4_2bi.html}}. This increases the number of effective noise pixels ($n_\mathrm{eff}$) for point sources, which scales like FWHM$^2$, while the limiting flux then scales as $\sqrt{n_\mathrm{eff}} \propto$ FWHM, so that the post-reactivation sensitivity is correspondingly decreased by a few hundredths of a mag. Furthermore, the W1 and W2 sensitivities have decreased slightly post-reactivation\footnote{\url{http://wise2.ipac.caltech.edu/docs/release/neowise/expsup/sec2_1e.html}, see Figure 1.}. The post-reactivation sensitivity decrease ranges from 0.05$-$0.12 mag in W1 and 0.15-0.26 mag in W2 when measured in yearly intervals during the NEOWISE mission. Combined, the increased PSF size and decreased sensitivity of post-reactivation data can account for the discrepancy between our achieved depths and the simplistic projection of 0.87 mag improvement. Confusion noise is an additional factor that would tend to reduce the depth improvement achieved relative to estimates based purely on reduced statistical pixel noise.

To measure the unWISE Catalog's reliability, we wish to compare against a highly complete Spitzer catalog. Therefore, in our reliability analysis, we compare against the entire S-COSMOS catalog without making any quality or flux cuts. We again require AllWISE \verb|w?snr| $\ge$ 5 and unWISE \verb|flux|/\verb|dflux| $\ge 5$ in the band under consideration. For unWISE we required \verb|flags_unwise| = 0, and that none of \verb|flags_info| bits 1, 6, 7 be set (see Table~\ref{tab:flags}). For AllWISE we required \verb|w?cc_map| = 0. We find that the unWISE Catalog has roughly comparable reliability to AllWISE until reaching sufficiently faint magnitudes that AllWISE no longer contains any sources in the sky region analyzed. In both W1 and W2, AllWISE appears to have very high reliability up until the point that it contains no more sources. We interpret this behavior as arising from the fact that AllWISE catalog construction and particularly its artifact flagging were engineered for high reliability, with the AllWISE Reject Table being used as a repository for lower confidence sources. The unWISE Catalog has no ``reject table'', and so its reliability rolls off toward faint magnitudes until reaching 0.62 (0.70) in W1 (W2), at which point there are no fainter $\ge 5\sigma$ sources in this sky region. The vast majority of AllWISE sources are detected at $\ge 5\sigma$ in W1, so that the W1 AllWISE sample reflected in the bottom left panel of Figure \ref{fig:completeness_reliability} can be considered roughly W1-selected. The unWISE sample in that same subplot is strictly W1-selected, making this a fair comparison. The W1 unWISE reliability is superior to the AllWISE reliability at the faintest AllWISE magnitudes, which makes sense given that the unWISE Catalog has benefited from a larger amount of W1 imaging. On the other hand, in W2, the AllWISE reliability appears to be slightly higher than the unWISE reliability at the faintest AllWISE magnitudes. We attribute this to the fact that AllWISE W2 detections are benefiting from W1 information (e.g., during the source identification and centroiding) as a result of being jointly fit across all bands, while the unWISE Catalog fits the two bands entirely independently. The relatively high AllWISE reliability around W2 $\sim$ 16 mag also coincides with AllWISE completeness which is much lower than that of the unWISE Catalog at the same magnitudes.

The depth of the unWISE Catalog varies across the sky, generally becoming deeper at higher ecliptic latitude and shallower at lower Galactic latitude.  Especially in W2, zodiacal light further reduces the depth at low ecliptic latitude.  The COSMOS field is at an ecliptic latitude of $\sim 10\degree$, so the COSMOS field corresponds to a relatively shallow region in unWISE.

\subsection{Astrometry}
\label{subsec:astrometry}

The simplest test of the unWISE Catalog astrometry is to compare it with Gaia  \citep{Gaia:2016}.  The comparison shows systematic offsets in declination of about -1 and 0 mas in right ascension ($\alpha$) and 28 and 36 mas in declination ($\delta$), in W1 and W2, respectively.  These offsets vary spatially over the sky, dominated by signals related to Galactic rotation; the rms scatter in the offset over the sky is 51 and 51 mas in $\alpha$ and 38 and 39 mas in $\delta$ in W1 and W2, respectively.  After removing these offsets over the sky, the rms scatter in $\alpha$ and $\delta$ for individual bright stars ($G < 15$) between Gaia and unWISE is 39 mas.

\begin{figure*}[htb]
\begin{center}\includegraphics[width=\textwidth]{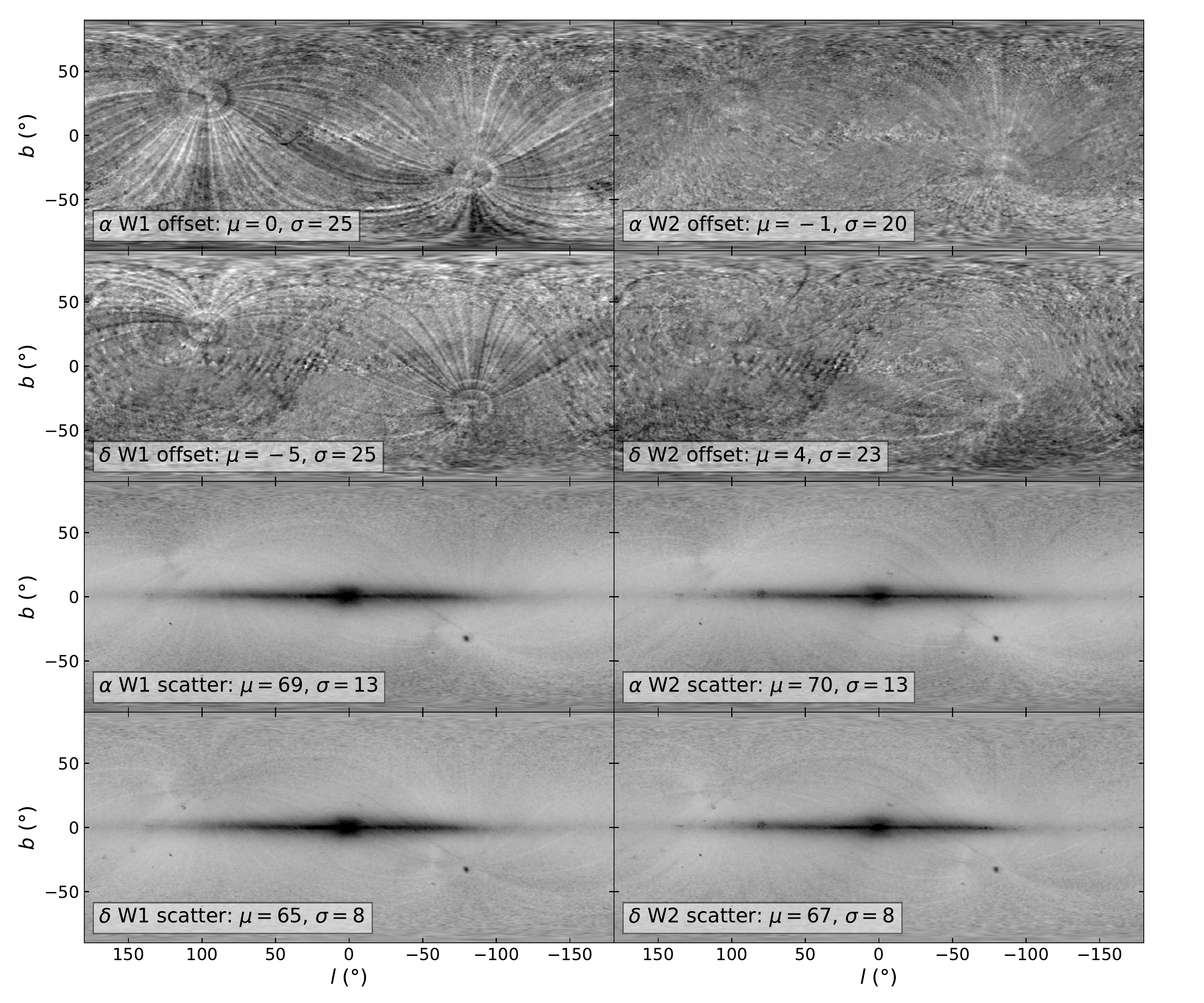}\end{center}
\caption{
\label{fig:astromaw}
Astrometry comparison between unWISE and AllWISE for bright, relatively isolated stars.  Each panel is a map of the sky; the rows shows the average difference in right ascension $\alpha$ and declination $\delta$, as well as the rms difference in $\alpha$ and $\delta$ over the sky.  All units are in milliarcseconds (mas).  The left column is W1, while the right column is W2.  The mean $\mu$ and standard deviation $\sigma$ of each map above $|b| = 15\degree$ is shown.  Agreement is excellent; typical correlated residuals are only 25 mas, less than a hundredth of a WISE pixel.  Individual bright stars agree in position to an rms difference of 70 mas.  Nevertheless, many WISE survey and processing related structures are evident; see text for details.  The color scale covers 200 mas linearly in all panels.
}
\end{figure*}

That said, a number of features appear in the comparison of the astrometry to Gaia unrelated to Galactic rotation.  To inspect these in more detail, we compare the unWISE astrometry to AllWISE.  In analyzing the unWISE coadds, we attempt to directly propagate the underlying WISE single-exposure astrometry forward without adjustment.  The unWISE Catalog astrometry should therefore agree quite well with the AllWISE astrometry.   Figure~\ref{fig:astromaw} shows the mean and root-mean-square (rms) difference between the measured unWISE and AllWISE right ascension (top) and declination (bottom) across the sky in W1 (left) and W2 (right) for bright stars ($W1 < 14$ or $W2 < 14$).  Overall, agreement is excellent.  There are spatially correlated differences of 25 mas in size, but this is only a hundredth of a WISE pixel.  Individual bright stars agree in position in AllWISE and unWISE with an rms difference of 70 mas.  We note that since AllWISE simultaneously fits the W1 and W2 bands, but W1 provides substantially more signal-to-noise for typical sources, Figure~\ref{fig:astromaw} primarily compares unWISE W1 and W2 to AllWISE W1.  In particular, differences between the W1 and W2 comparisons stem from differences in the unWISE W1 and W2 astrometry; the AllWISE astrometry is identical.

Despite this great overall agreement, Figure~\ref{fig:astromaw} shows interesting structures in the differences between the unWISE and AllWISE astrometry.  Most prominently, the differences in $\alpha$ and $\delta$ show stripes along lines of constant ecliptic longitude.  This is due to a difference between the definitions of the ``center'' of the PSF in unWISE and AllWISE.  AllWISE adopted a PSF with a flux-weighted centroid slightly offset from the origin, while unWISE uses a PSF with a centroid at the origin \citep{tr_neo2}.  However, unWISE adopts the published WISE WCS solutions without modification, leading unWISE to report slightly different world coordinates for sources than AllWISE.  This difference is most pronounced in locations where, due to Moon avoidance maneuvers, more observations have been made in one WISE scan direction than in the other, leading to structures at constant ecliptic longitude.  The larger amplitude of the striping in W1 than W2 is due to the greater PSF asymmetry in W1 than W2 \citep{tr_neo2}.

Another source of low-level disagreement between unWISE Catalog positions and AllWISE positions are the ``MFPRex'' astrometric corrections
\citep[][$\S$V.2]{Cutri:2013} implemented in the AllWISE pipeline but not reflected in the single-exposure WCS used by unWISE. These corrections presumably are the source of the coherent small scale features in Figure~\ref{fig:astromaw}, most prominent near the Galactic Center in the $\delta$ offset map (second row).  These corrections additionally account for the proper motions of the stars in the astrometric reference catalog, leading to some small differences between AllWISE and unWISE on large angular scales.  Comparison to Gaia confirms that the AllWISE astrometry is superior to unWISE, owing to these corrections, so users seeking the best possible astrometry are encouraged to use AllWISE until the unWISE astrometry can be improved (\textsection\ref{subsec:astrometrylimits}).

The $\delta$ offset maps (second row of Figure~\ref{fig:astromaw}) show significantly smaller residuals in the celestial south than in the north; we do not understand the cause of this north-south discrepancy.

The rms differences in $\alpha$ and $\delta$ are very uniform over the high-latitude sky.  In the inner Galaxy, however, they increase dramatically, presumably owing to the more aggressive unWISE identification of stars in the wings of neighboring stars and the corresponding impact on the positions of the stars and their neighbors.  Similarly, the Magellanic Clouds and M31 show enhanced rms.  Two small regions of slightly enhanced residuals and scatter in $\delta$ fall near (113\degree, 17\degree) and (148\degree, -18\degree) and are not understood, though the modeling in these regions looks accurate, leading us to suspect problems with the WCS.

\subsection{Photometry}
\label{subsec:photometry}
\begin{figure*}[htb]
\begin{center}\includegraphics[width=\textwidth]{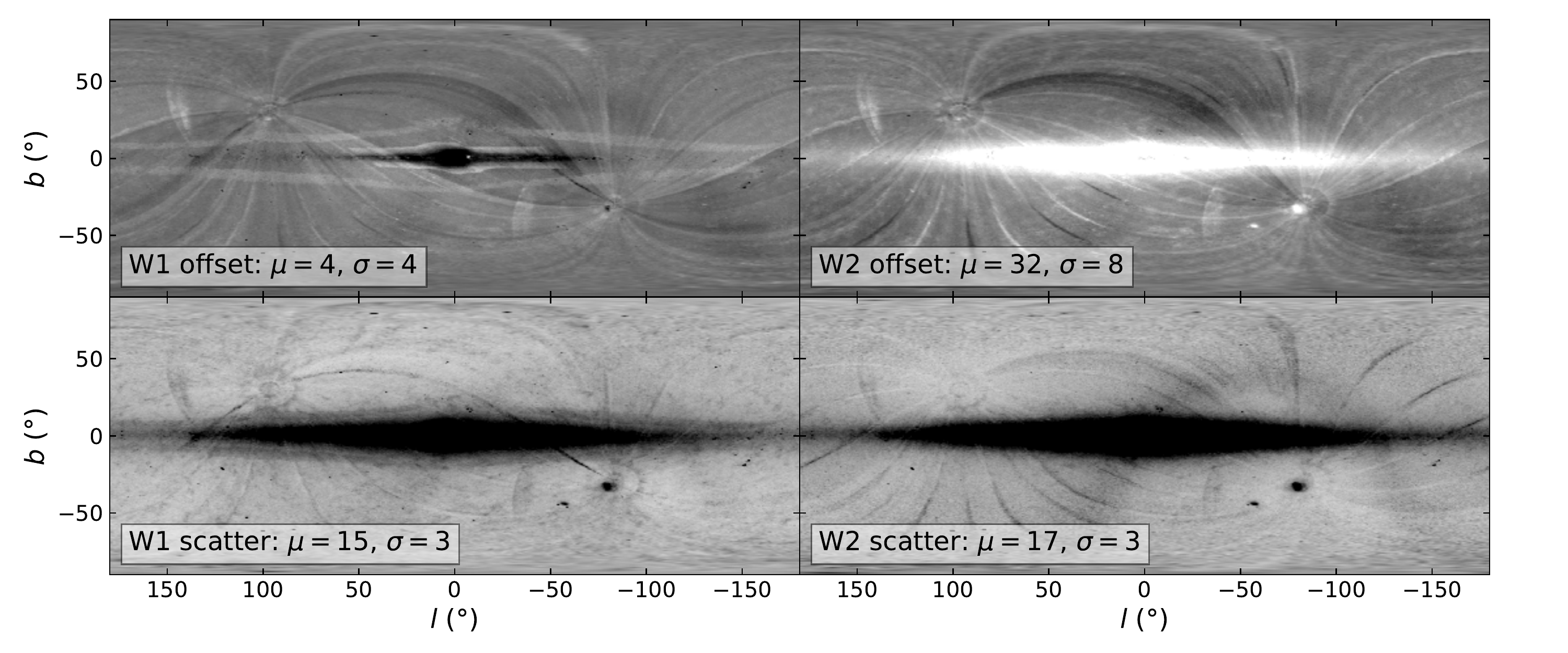}\end{center}
\caption{
\label{fig:photomaw}
Photometry comparison between unWISE and AllWISE for bright (W1 or W2 brighter than 14th mag), relatively isolated stars.  Each panel is a map of the sky; the first row shows the average difference in magnitude between unWISE and AllWISE, while the second row shows the rms scatter in the magnitude differences.  The left column shows W1, while the right column shows W2.  The mean $\mu$ and standard deviation $\sigma$ of each map above $|b| = 15\degree$ is shown, with units of mmag.  Agreement is excellent, with percent-level offsets and uniformity, and scatter of $1\%$, though many WISE survey and processing related structures are evident; see text for details.  The color scale covers 100 mmag linearly in the top two panels and 50 mmag linearly in the bottom two panels.
}
\end{figure*}

To assess the accuracy of the unWISE photometry, we compare it with AllWISE photometry in Figure~\ref{fig:photomaw}.  Agreement is excellent, with overall offsets of 7 and 35 mmag in W1 and W2 and sub-percent level fluctuations over the high-latitude sky.  The rms difference between unWISE and AllWISE photometry of bright stars is 15 mmag in W1 and 17 mmag in W2.

However, a number of structures are evident in the comparison.  Most prominently, the comparison features a number of streaks running between the ecliptic poles at constant ecliptic longitude.  These appear to stem from the AllWISE photometry, as they correlate with the moon-avoidance maneuvers made during the pre-hibernation phase of the WISE mission; the post-reactivation WISE imaging has now covered the sky more uniformly.  Structures are apparent at $(l, b) = (145\degree, 45\degree)$ and $(-40\degree, -40\degree)$, which stem from the WISE spacecraft's dumping angular momentum using its magnetic torque rods  \citep[][IV.2.c.i.]{Cutri:2013}.  A dark band of higher-than-average scatter is apparent in the W2 rms map at $\delta \approx -30\degree$, due to the South Atlantic Anomaly.  The improved cosmic-ray rejection made possible by the many NEOWISE epochs \citep{fulldepth_neo2} allows the unWISE photometry to improve on the AllWISE photometry in this region.  Very crowded regions and regions with significant nebulosity appear as areas of poor AllWISE-unWISE agreement: for example, the Galactic plane, the Large and Small Magellanic Clouds, large globular clusters, and the Orion nebula.  The W1 offset map additionally shows some peculiar regions around the Galactic plane and inner bulge where stars are preferentially a couple hundredths brighter in unWISE than in AllWISE.  We do not understand these features, but comparison with 2MASS indicates that they are present in AllWISE and absent from unWISE.

\begin{figure}[htb]
\begin{center}\includegraphics[width=\columnwidth]{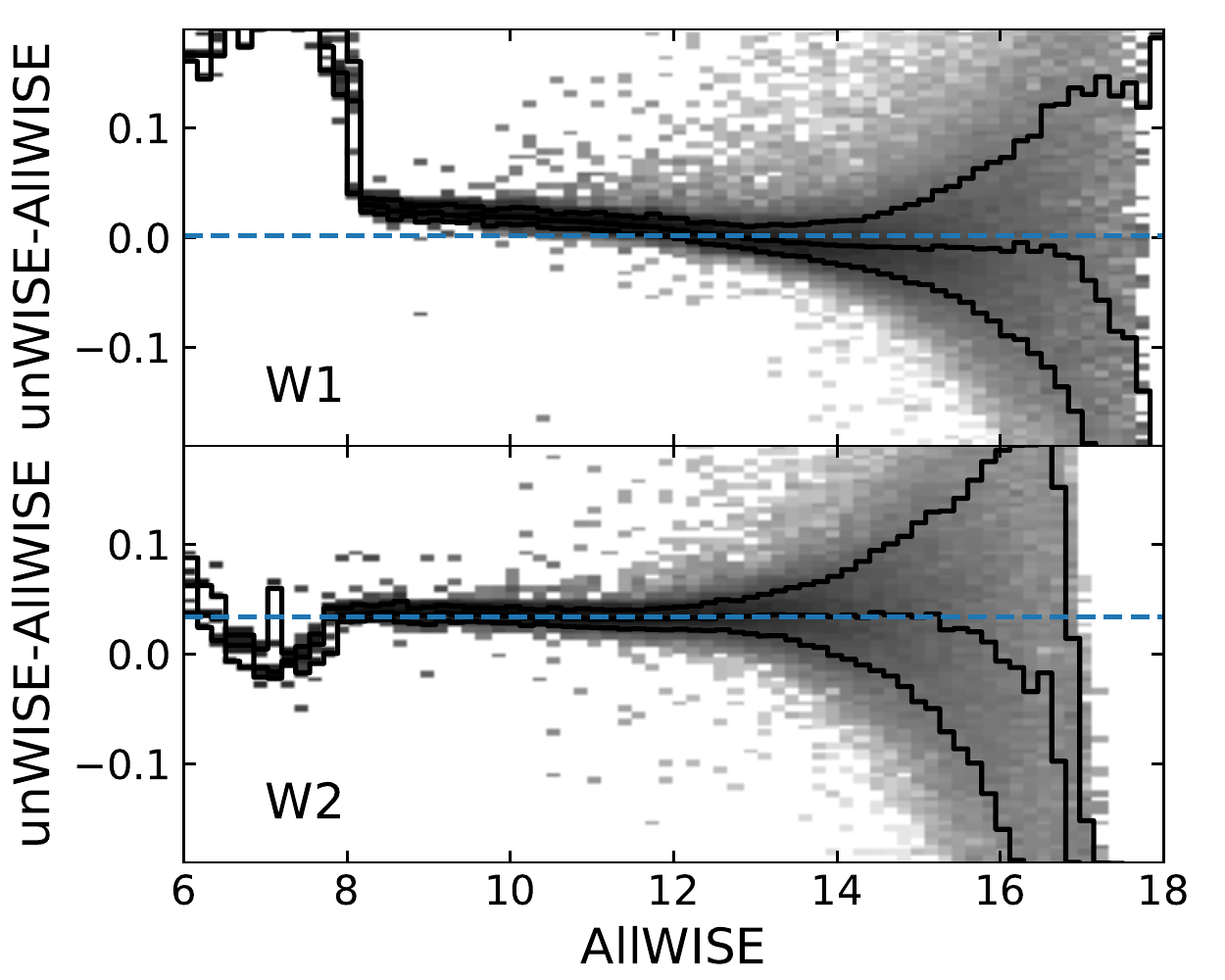}\end{center}
\caption{
\label{fig:dphotomvmag}
Photometry comparison between unWISE and AllWISE as a function of magnitude.  The grayscale shows the number of stars at each unWISE-AllWISE magnitude difference as a function of the magnitude of the stars.  The solid lines show the 16th, 50th, and 84th percentiles of the distribution at each magnitude.  There is good agreement between AllWISE and unWISE at bright magnitudes fainter than the WISE saturation limit of 8, with about 15 mmag scatter.  W2 shows excellent linearity, but in W1 there is a trend of 40 mmag from 8th to 16th mag, with unWISE being fainter than AllWISE for bright stars and brighter than unWISE for faint stars.
}
\end{figure}

Figure~\ref{fig:photomaw} addresses the spatial uniformity of the photometry of bright stars, but is insensitive to the catalog accuracy of faint stars.  Figure~\ref{fig:dphotomvmag} shows the differences between unWISE and AllWISE photometry as a function of magnitude for point sources identified by Gaia in a 25 square degree region around the COSMOS field.  Agreement is again good; for bright stars the rms difference is a few hundredths, as anticipated in Figure~\ref{fig:photomaw}, and near the AllWISE faint limit the uncertainties increase to the expected $\approx 0.2$ magnitudes.  In W2, the unWISE and AllWISE fluxes have a tight linear relationship between the saturation limit at about 8th magnitude to the faint limit at about 16th magnitude.  However, in W1 there is a clear trend in the magnitude difference with magnitude; unWISE sources are 0.03 mag fainter than AllWISE sources at 8th mag, while they are 0.01 mag brighter than AllWISE sources at 17th mag.  It is unclear where this nonlinearity comes from; one hint is that the measured sizes of point sources likewise depends on magnitude, albeit in both W1 \emph{and W2}; see \textsection\ref{subsec:nonlinearity} for further discussion.  At about 8th mag in both W1 and W2, the unWISE magnitudes depart sharply from the AllWISE magnitudes.  This is due to the onset of saturation.  Due to the way the unWISE patches the saturated cores of stars in the construction of the deep unWISE coadds \citep{lang_unwise_coadds}, the inner $7\times7$ pixel regions of saturated stars are unreliable.  Flux estimates of saturated stars are then made entirely on the basis of the wings of the PSF in the unWISE Catalog, outside the $7\times7$ pixel region where roughly 90\% of the flux resides.  The onset of saturation is tracked in the unWISE Catalog columns \verb|flags_unwise| and \verb|qf|, allowing easy identification of saturated sources (\textsection\ref{sec:release}).

The unWISE Catalog absolute photometric calibration derives from the photometric calibration of the unWISE coadds \citep{fulldepth_neo1}, which is tied to the original WISE zero points through aperture fluxes in a 27.5\arcsec\ radius.  The unWISE Catalog fluxes are defined in the context of a PSF that is normalized to unity in a $19\times19$ pixel box, corresponding to 52.25\arcsec.  Meanwhile the AllWISE PSFs are normalized to unity in a 220\arcsec\ box.  Because the PSF is extremely stable, these different conventions lead one to expect slight offsets ($\sim$30 mmag) in the absolute calibration of unWISE and AllWISE, consistent with Figure~\ref{fig:photomaw}.  We recommend subtracting 4 mmag and 32 mmag in W1 and W2 from the unWISE magnitudes to better match the AllWISE absolute flux calibration, with the caveat that the W1 correction is particularly uncertainty because of the W1 nonlinearity (Figure~\ref{fig:dphotomvmag}).

\subsection{Example Uses}
\label{subsection:examples}

The unWISE Catalog should prove a valuable resource for a range of astronomical applications.  This section presents simple examples of potential uses in Galactic, extragalactic, and high-redshift science.

\subsubsection{Galactic}
\label{subsec:galacticcmd}

\begin{figure*}[htb]
\begin{center}\includegraphics[width=\textwidth]{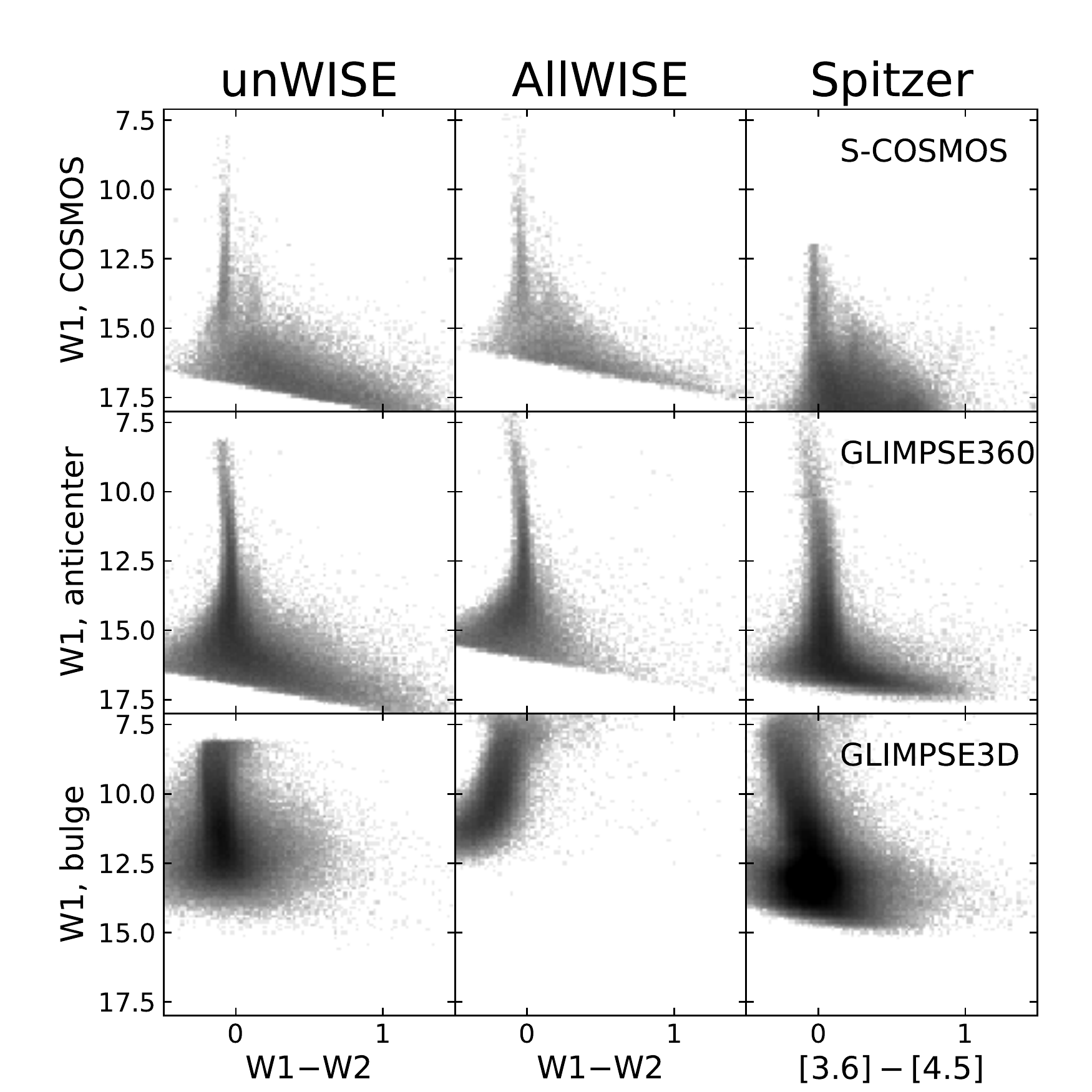}\end{center}
\caption{
\label{fig:cmds}
Color-magnitude diagram of $5\sigma$ sources from unWISE, AllWISE, and Spitzer in three different fields: the high-latitude COSMOS field, the Galactic anticenter, and the Galactic bulge.  The greater depth of the unWISE catalog relative to AllWISE is immediately apparent.  In relatively uncrowded high-latitude fields, unWISE extends roughly 0.7~mag fainter than AllWISE, as expected from \textsection\ref{subsec:completeness}.  This continues to be the case in the Galactic anticenter.  In the Galactic bulge, both the unWISE and AllWISE color-magnitude diagrams clearly suffer from crowding, with broad color distributions even at bright magnitudes.  Comparison with much higher resolution Spitzer observations from S-COSMOS, GLIMPSE360, and GLIMPSE3D show good agreement with the unWISE magnitudes.  The deep Spitzer observations on the COSMOS field are naturally much deeper than unWISE, but unWISE is competitive with GLIMPSE in depth in the outer Galaxy, while providing a tighter stellar locus.  In the inner Galaxy, the higher-resolution GLIMPSE observations allow for deeper catalogs.}
\end{figure*}

A wide variety of Galactic science relies on WISE infrared fluxes, ranging from nearby stars \citep{kirkpatrick2011} to dust-extinguished stars throughout the Galaxy \citep{Schlafly:2016, Schlafly:2017}.  Figure~\ref{fig:cmds} illustrates the improvement of the unWISE fluxes relative to AllWISE in three fields of different source densities: the high-latitude COSMOS field (top row), the Galactic anticenter (middle row), and the Galactic bulge (bottom row).  The figure also shows high-resolution data from Spitzer, taken from the ultra deep S-COSMOS program on the COSMOS field \citep{Sanders:2007}, from GLIMPSE360 in the Galactic anticenter, and from GLIMPSE3D in the Galactic bulge \citep{Benjamin:2003, Churchwell:2009}.  In each case, the greater depth of the unWISE imaging (left) than the AllWISE imaging (center) is apparent; the sources in the color-magnitude diagrams extend roughly 0.7~mag fainter in the unWISE diagrams than the AllWISE diagrams.

In the COSMOS field and toward the Galactic anticenter, the unWISE measurements are very similar to the AllWISE measurements and extend them naturally to fainter magnitudes.  In the bulge, however, there is a pronounced difference between the unWISE and AllWISE color-magnitude diagrams; the typical star in AllWISE becomes bluer at fainter magnitudes, while it becomes redder in unWISE.  The higher resolution Spitzer observations better match the unWISE measurements than the AllWISE measurements, giving us confidence that unWISE is improving on AllWISE in dense regions like the bulge.  That said, in these regions the Spitzer data are substantially superior to the unWISE data---unsurprisingly, given the extreme crowding and $\sim4\times$ better resolution of Spitzer.

\subsubsection{Extragalactic}
\label{subsec:extragalacticnz}

\begin{figure}[htb]
\begin{center}\includegraphics[width=\columnwidth]{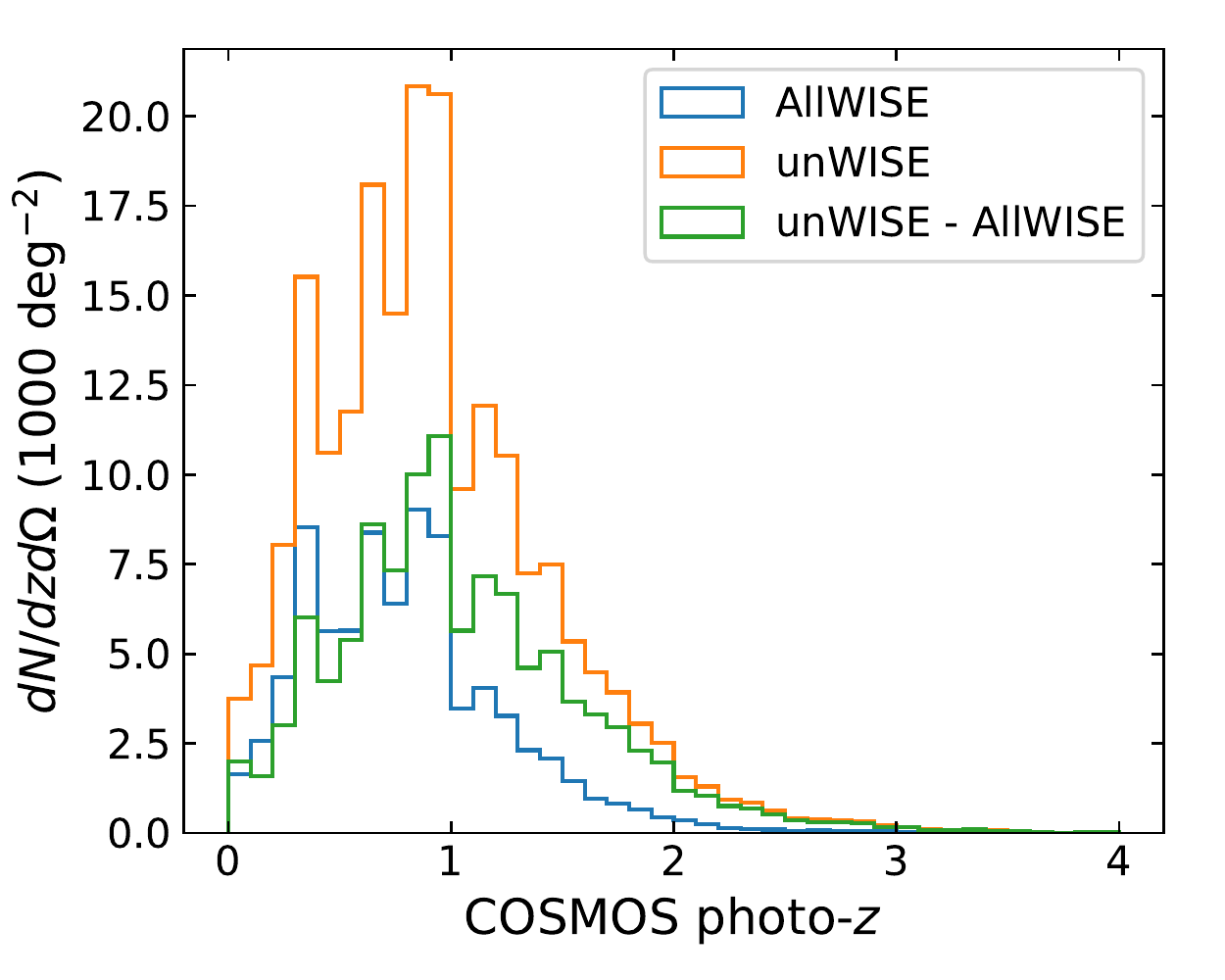}\end{center}
\caption{
\label{fig:cosmosnz}
Redshift distribution of galaxies in the COSMOS field detected in AllWISE (blue), unWISE (orange), and the difference (green).  The unWISE catalog roughly doubles the number of galaxies detected with $z<1$, while tripling it at $z>1$.   Extrapolating to the entire sky, the unWISE catalog should contain $>500$ million galaxies broadly distributed over $0 < z < 2$. 
}
\end{figure}

At $z=0$, the WISE W1 and W2 bands sample a steeply falling portion of the typical early-type galaxy's spectral energy distribution.  At higher redshifts, more and more of a galaxy's light redshifts into the WISE bands.  This makes WISE an effective tool for detecting galaxies at redshifts $0 < z < 2$.  To illustrate this, Figure~\ref{fig:cosmosnz} shows the redshift distribution of galaxies detected by AllWISE and unWISE in the COSMOS field \citep{scosmos}, with redshifts taken from the photometric redshift catalog of \citet{Laigle:2016}.  WISE sources were matched to COSMOS sources with a match radius of 2.75\arcsec, considering only COSMOS sources with Spitzer 3.6\mum\ or 4.5\mum\ fluxes bright enough that the sources could conceivably be detected in WISE.  The resulting redshift distribution peaks at $z\approx 1$.  The bulk of the distribution falls between $0 < z < 2$, with a tail to higher redshifts.

Figure~\ref{fig:cosmosnz} further indicates that unWISE roughly doubles the number of $0 < z < 1$ galaxies detected, while increasing the number of galaxies with $1 < z < 2$ by a factor of three or more.  Extrapolated over the whole sky, the unWISE catalog should contain $>500\times10^6$ galaxies with $0 < z < 2$.  Table~\ref{tab:nz} summarizes the number densities of sources of different types and redshifts in the COSMOS field for AllWISE and unWISE.

\begin{deluxetable}{c|cc|ccccc}
\tablecaption{Number density of objects at different redshifts}
\tablehead{
\colhead{} & \multicolumn{2}{c}{stars} & \multicolumn{5}{c}{$z$ range} \\
\colhead{Catalog} & \colhead{Gaia} & \colhead{$\neg$Gaia} & \colhead{0, 0.5} & \colhead{0.5, 1} & \colhead{1, 1.5} & \colhead{1.5, 2} & \colhead{$>2$}
}
\startdata
\label{tab:nz}
AllWISE & 2977 & 411 & 2151 & 3760 & 1501 & 411 & 126 \\
unWISE & 4562 & 1297 & 3941 & 8457 & 4598 & 1876 & 773 \\
\enddata
\tablecomments{Number of objects per square degree for stars and galaxies of different redshifts, based on comparison to objects with photometric redshifts in COSMOS \citep{Laigle:2016}.  unWISE increases the number of galaxies detected by a factor of 2--4.  Stars are marked as having been identified by Gaia (Gaia), or not ($\neg$Gaia).  Note that counts are given only for objects matching objects in \citep{Laigle:2016}, but roughly 6\% of unWISE objects have no match, primarily due to masked regions near bright stars in COSMOS and large galaxies split into multiple PSF components in unWISE.}
\end{deluxetable}

For extragalactic purposes, the presence of stars in the catalog can be a nuisance.  Often, however, these stars can be identified by their pointlike morphology in Gaia imaging, which like WISE, is available for the entire sky.  Table~\ref{tab:nz} indicates the number density of stars detected by Gaia, and the number density not detected by Gaia, usually due to faint magnitudes and red colors.

\begin{figure}[htb]
\begin{center}\includegraphics[width=\columnwidth]{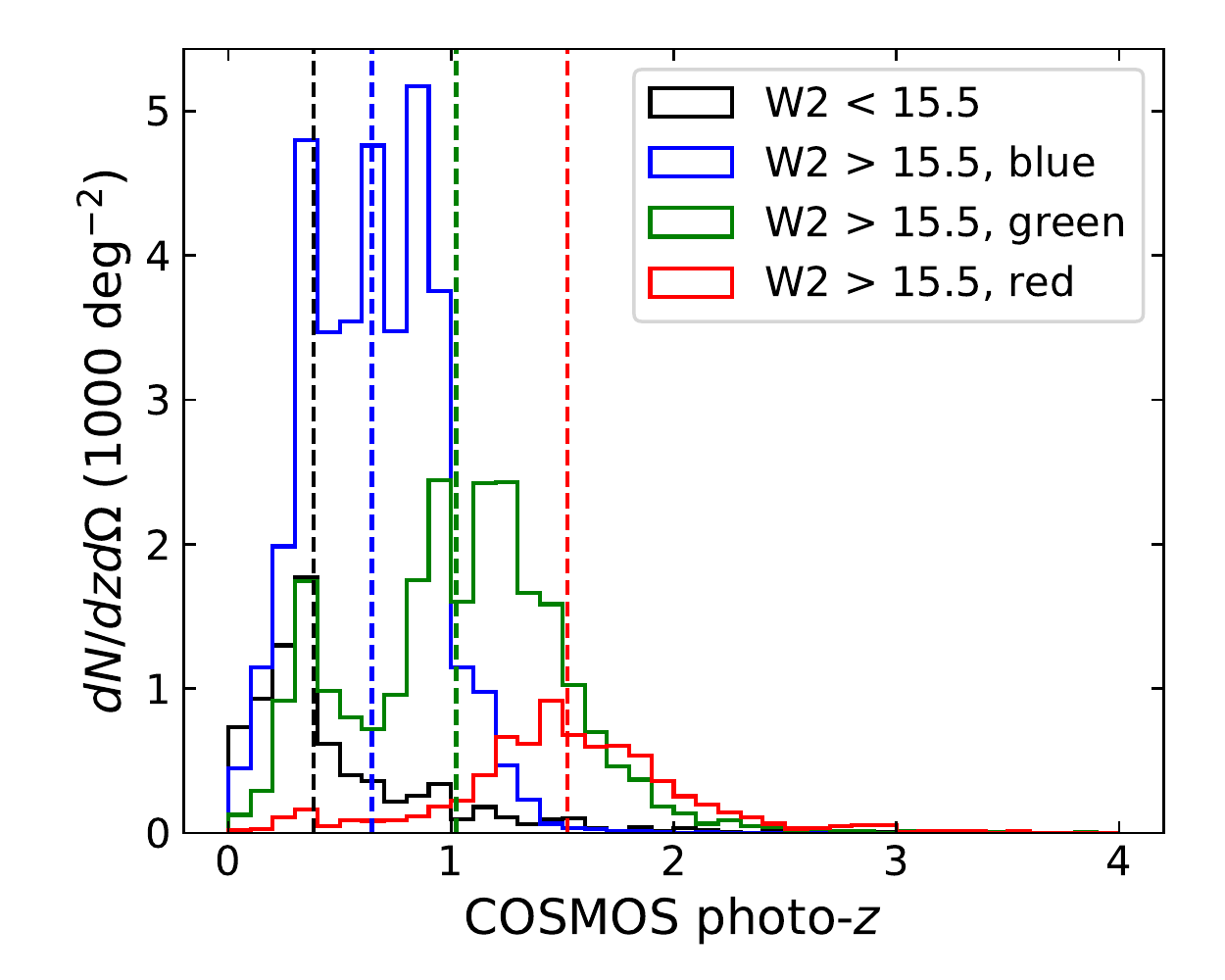}\end{center}
\caption{
\label{fig:cosmosnz2}
Redshift distribution of galaxies in the COSMOS field detected in unWISE, satisfying four different color and magnitude selections.  Judicious cuts on galaxies' WISE colors can produce samples with mean redshifts ranging from 0.4 to 1.5.  Vertical lines give the mean redshifts of the different selections.
}
\end{figure}

From the unWISE Catalog alone, the only information available about a typical galaxy is its flux in the W1 and W2 bands.  This makes efforts to estimate a galaxy's redshift from its unWISE Catalog entry challenging.  Nevertheless, there is a good correlation between the WISE color of a galaxy and its redshift.  Figure~\ref{fig:cosmosnz2} shows the redshift distribution of unWISE Catalog galaxies in the COSMOS field satisfying four simple color cuts.  The galaxies passing these cuts have mean redshifts steadily increasing from $z=0.4$ to $z=1.5$, with a typical rms of 0.4, as detailed in Table~\ref{tab:nz2}.

\begin{deluxetable}{ccccc}
\tablecaption{Color cuts for different WISE galaxy samples}
\tablehead{
\colhead{W2} & \colhead{W1$-$W2$ > x$} & \colhead{W1$-$W2$ < x$} & \colhead{$\bar{z}$} & \colhead{$\delta z$}
}
\startdata
\label{tab:nz2}
$<15.5$ & & & 0.4 & 0.3 \\
$> 15.5$ & & $(17-\mathrm{W2})/4+0.3$ & 0.6 & 0.3 \\
$> 15.5$ & $(17-\mathrm{W2})/4 + 0.3$ & $(17-\mathrm{W2})/4 + 0.8$ & 1.0 & 0.4 \\
$> 15.5$ & $(17 - \mathrm{W2})/4 + 0.8$ & & 1.5 & 0.4 \\
\enddata
\tablecomments{Color and magnitude cuts for selecting galaxies of different redshifts, together with the mean redshift $\bar{z}$ and the width of the redshift distribution $\delta z$ for the selections, as measured by matching to objects with photometric redshifts on the COSMOS field \citep{Laigle:2016}.}
\end{deluxetable}

\subsubsection{High-redshift}
\label{subsec:highredshiftqso}

Mid-infrared colors provide an efficient means of selecting quasars, making them effective for detecting objects at high redshifts \citep{Wang:2016}.  By providing deep mid-infrared photometry, the unWISE Catalog should prove valuable in searches for the highest redshift quasars.

Consistent with this expectation, the unWISE catalog contains detections of  more $z > 5$ quasars than AllWISE.  Among the 453 quasars currently known (Ross \& Cross, in prep.), 268 are detected in W1 and 183 are detected in W2 in AllWISE, where ``detection'' means that the catalog contains a source within 2.75\arcsec\ of the QSO with $>5\sigma$ significance.  Meanwhile, 355 and 307 are detected in unWISE in W1 and W2, respectively.  Roughly half of all high-redshift quasars formerly undetected in WISE now have secure detections.

\section{Limitations and Future Directions}
\label{sec:limitations}

The unWISE Catalog is a first attempt to measure the fluxes and positions of all of the sources in the unWISE coadds.  During the construction of the catalog, a number of issues were identified that could be addressed in future unWISE catalogs.  We list some limitations of the catalog here, and in many cases describe improvements that could eliminate these limitations in future releases.

\subsection{Non-linearity}
\label{subsec:nonlinearity}
Comparison of AllWISE and unWISE magnitudes reveals a slight non-linearity in W1, as evident in Figure~\ref{fig:dphotomvmag}.  It is not clear what the source of this trend is.  Variation in the non-linearity of the WISE detector pixels between the mission phases could account for some of the effect.  Inspection of the shape of the PSF as a function of magnitude (as measured by \texttt{spread\_model}) shows a significant trend in both W1 and W2, large enough to account for the trend in W1.  However, because the shape-dependent trend is present in both W1 and W2, the obvious avenues for eliminating it introduce a significant trend between AllWISE and unWISE in W2.  No correction for non-linearity is made in the unWISE catalog, incurring systematic uncertainties of $\approx 0.02$ mag.  More work is needed to identify the source of the non-linearity and to develop effective mitigation strategies.

\subsection{Saturation}
\label{subsec:saturation}
The performance of the unWISE Catalog rapidly degrades at the onset of saturation, as indicated in Figure~\ref{fig:dphotomvmag}.  The unWISE coadd images (\textsection{\ref{sec:unwise}}) use a $7\times7$ Lanczos kernel to resample the native WISE images onto the unWISE coadds.  Before resampling, any masked pixels are first ``patched'' by replacing the masked value with the mean of the surrounding pixels.  This procedure tends to make the peaks of stars flatter than they would otherwise have been, changing the shape of the PSF.  This changed shape is then propagated out to the full $7\times7$ pixel surrounding neighborhood, albeit with decreasing influence toward the edges.

Consistent with this, saturated stars show significantly worse residuals than unsaturated stars in the unWISE imaging, even outside the saturated center.  To mitigate this effect, all pixels within a three pixel radius of a potentially saturated pixel in unWISE are masked in the analysis.  Because roughly $90\%$ of the flux of a star lands within 3 pixels of a star's center, the unWISE catalog fluxes of these bright stars are highly uncertain and dependent on the amount of flux in the wings of the PSF.

Significant improvement here could come from improving the ``patching'' process for saturated pixels in the unWISE coadds by incorporating knowledge of the WISE PSF.

\subsection{Sky Subtraction}
\label{subsec:skysubtractionlimitation}

The \texttt{crowdsource} sky subtraction analysis removes the median residual in each $19\times19$ pixel region of the image during each iteration of its fitting process.  In very dense regions like the Galactic bulge this process tends to be biased high, and many potential sources are not discovered and analyzed.

The sky could instead be fit as an additional set of linear parameters in the main \texttt{crowdsource} optimization.  In the DECam Plane Survey \citep{Schlafly:2018}, \texttt{crowdsource} was run in a mode in which a single global sky was simultaneously optimized, but in unWISE we found that this tended to propagate small residuals around bright stars into global sky errors that interacted with the median sky filtering to slow convergence.  The locality of the sky fit could be preserved with simultaneous fitting of the sky by adopting a small-scale cardinal basis spline approach to the sky modeling.  This approach would require somewhat more memory than the existing analysis, but should improve the speed of convergence and the accuracy of the sky fits.

\subsection{Inconsistency of W1 \& W2 Modeling}
\label{subsec:w1w2modeling}

Sources in the unWISE Catalog are modeled completely independently in different bands, in contrast to AllWISE.  This means that images that are modeled with three stars in W1 may be modeled with two stars in W2, confusing the linking of W1 and W2 catalogs into multiband object lists.  Similarly, a single, isolated object will have slightly different positions in its W1 and W2 analysis in the unWISE Catalog, again in contrast to AllWISE.

The AllWISE approach has many advantages over the approach taken for unWISE; the primary motivation for the unWISE approach was computational and algorithmic convenience.  Future releases could straightforwardly adopt an AllWISE-like simultaneous modeling scheme.  Nevertheless, color-magnitude diagrams like Figure~\ref{fig:cmds} indicate that the negative effects of the inconsistent modeling are not large for typical stars.

\subsection{No source motions or variability}
\label{subsec:staticsky}

The unWISE Catalog is strictly a static-sky catalog; it does not fit for the motions of sources or for their variability.  Time-resolved unWISE coadds \citep{tr_neo2} preserve almost all of the information present in the WISE data about the motion of objects outside of the Solar System.  In principle, analysis of these images could recover proper motions for many stars.  Like \textsection\ref{subsec:w1w2modeling}, this would require simultaneously modeling several images, but beyond this, the generally very small, subpixel motions would be naturally accommodated in the sparse linear algebra analysis at the core of \texttt{crowdsource}.

Similarly, the unWISE Catalog contains no variability information; this has likewise been lost in the construction of the deep static-sky unWISE coadds we have analyzed.  Variability on time scales of 0.5--8 years, however, would be accessible through analysis of the time-resolved unWISE coadds.

\subsection{Astrometry}
\label{subsec:astrometrylimits}

The unWISE coadds adopt the world-coordinate system of the underlying WISE single-exposure images without modification.  The WISE single-exposure world-coordinate system is correct in the context of a slightly asymmetric PSF model\footnote{See \url{http://wise2.ipac.caltech.edu/docs/release/neowise/expsup/sec3\_2.html\#astrom} and $\S$4.1 of \cite{tr_neo2} for in-depth discussion of this single-exposure PSF model asymmetry.} that is different from the PSF model adopted for the unWISE Catalog.  The resulting inconsistency leads to the $\approx 25$ milliarcsecond residuals throughout the upper two rows of Figure~\ref{fig:astromaw}.

The unWISE coadds were made from single-epoch images that did not include MFPRex WCS improvements.  Inclusion of these improvements and correction for the different AllWISE and unWISE PSF conventions would remove the dominant sources of coherent astrometric residuals in unWISE.

Another option for improving the unWISE Catalog's astrometric accuracy would be to recalibrate either the single-exposure WISE image astrometry or the unWISE coadd astrometry, using a procedure similar to that described in \cite{tr_neo2}.  The release of unprecedentedly accurate astrometry and proper motions from Gaia should enable substantial improvements to the native WISE single-exposure astrometry (currently tied to 2MASS, in some cases using UCAC4 proper motions).

\subsection{Bright Stars}
\label{subsec:brightstars}

The WISE PSF extends very far from the source; flux is readily detected 2\degree\ from the centers of the brightest stars.  The PSF used for the unWISE analysis extends ``only'' $\sim 150$ pixels ($\sim 0.1\degree$) from the centers of stars.  Moreover, the wings of the unWISE PSF are very uncertain.  Accordingly, there are significant residuals $\approx 0.1\degree$ from very bright stars that may present as spurious sources in the unWISE Catalog.  Closer to the center of extremely bright stars, large saturated regions likewise occasionally lead to the generation of spurious sources in the catalog.

Diffraction spikes around very bright stars pose an additional challenge.  The interpolation process from the original images onto new tile centers makes the detailed path a diffraction spike should follow in the coadd images subtle.  Far from the center of a star, the unWISE PSF diffraction spikes tend to fall slightly off the true diffraction spikes, leading to significant residuals and affecting the flux of stars in the vicinity.  Spurious sources may also be detected in these locations.

The unWISE images contain a number of flags indicating pixels affected by bright stars; these are included in the unWISE catalog in the \texttt{flags\_unwise} column (\textsection\ref{sec:release}).  Improved modeling of the wings of very bright stars is also possible, but would require significant effort while improving the analysis in only a tiny fraction of the sky.

\subsection{Extended Sources}
\label{subsec:extended}
All sources in the unWISE analysis are modeled as if they were point sources.  Fluxes and locations of extended sources will be correspondingly biased and sub-optimal.  However, the broad WISE PSF of 6\arcsec\ FWHM means that most extended sources can be treated as point sources.  For example, a 1\arcsec\ FWHM galaxy would only broaden the PSF by 1.4\%, corresponding to a flux estimate 1.4\% too small in the unWISE catalog.

Some galaxies, however, are much larger, and are readily resolved by WISE.  Absent intervention, \texttt{crowdsource} will split these galaxies into several point-source components.  Many of these large galaxies have already been identified in other surveys, however, making it easy to modify the behavior of \texttt{crowdsource} in these regions.  The unWISE mask images include a bit indicating that a pixel is contained in a large galaxy, as tabulated in the HyperLeda catalog \citep{Makarov:2014}.  In such pixels, \texttt{crowdsource} rejects candidate new sources if they significantly overlap with a neighboring source.

The effectiveness of this mitigation strategy depends on the underlying catalog of large galaxies.  The HyperLeda catalog is rather heterogeneous; over the Sloan Digital Sky Survey \citep{York:2000} footprint it contains many more large galaxies than in the southern sky, for instance.  The unWISE catalog will correspondingly spuriously split varying numbers of large galaxies into multiple point sources depending on sky location.

This strategy discourages \texttt{crowdsounce} from splitting large galaxies, but significant residuals are still left on the images around large galaxies because the point source model is a poor fit.  As part of the optimization process, \texttt{crowdsource} may still decide that the image is best modeled by slowly moving nearby stars into the large galaxy to reduce these residuals, leading to occasional splitting even of large galaxies identified in HyperLeda.

Explicitly modeling all of the sources as potentially resolved galaxies or unresolved stars would be the best approach, but presents substantial conceptual, algorithmic, and computational challenges.  Alternatively, full galaxy fitting could be enabled exclusively for objects marked in external catalogs as being resolved galaxies; this would improve the fits of these galaxies and their neighbors.  Galaxy models tend to be less well approximated by linear models, however, so their optimization does not fit neatly into the \texttt{crowdsource} framework.

\subsection{Ecliptic Poles}
\label{subsec:eclipticpoles}

The unWISE coadd images are qualitatively different at high ecliptic latitudes than at low ecliptic latitudes.  At low ecliptic latitudes, the typical number of WISE observations contributing to a given part of the sky in the unWISE coadds is  $\sim 120$ per band.  At the ecliptic poles, however, the number is greater than 23000,  $\sim 200\times$ larger.  Moreover, due to the WISE scan strategy, at low ecliptic latitudes, the position angle of the WISE focal plane on the sky is nearly constant, while the ecliptic poles were observed at all position angles.

The huge number of WISE observations at the ecliptic pole potentially make for images $\sim 3$ magnitudes deeper than typical at low latitudes.  However, at this depth, large-scale residuals in the unWISE coadd sky become significant.  Absent intervention, the \texttt{crowdsource} analysis would incorrectly attempt to explain what remains of these residuals after sky subtraction with many sources.  To mitigate this, a flux uncertainty floor was added to the unWISE uncertainty images such that the $5\sigma$ sources correspond to at most $\sim 19.8$ mag in W1 and $\sim 18.2$ mag in W2.  These are roughly 1.5 mag fainter than the typical depths of the unWISE coadd images (\textsection\ref{subsec:completeness}), but much brighter than the nominal depth achievable at the ecliptic poles.

The varying position angle of the WISE images taken near the ecliptic pole leads the PSF there to be different from the PSF at lower ecliptic latitudes.  As discussed in \textsection\ref{subsec:psf}, the \texttt{crowdsource} analysis uses a dense grid of rotated and azimuthally smoothed PSF models to capture much of the PSF variation in the unWISE coadds in these areas.  However, our flagging of diffraction spikes and ghosts in the unWISE coadd masks was not modified to account for the azimuthal smoothing of the PSF near the ecliptic poles.  Accordingly, these features in the unWISE masks near the ecliptic pole are narrower than they should be.

This problem primarily affects regions within about 5\degree\ of the ecliptic poles,  $\sim0.4$\% of the footprint.  Moreover, because sharp features like diffraction spikes spread over a wider area due to this effect, their amplitude is lower and their masking is less important.  Directly on the pole, for instance, it is not clear if a diffraction spike mask would be desired; the diffraction spikes have been fully blurred out into the wings of the PSF.

\subsection{Nebulosity}
\label{subsec:nebulositylimitations}

The \texttt{crowdsource} analysis identifies regions believed to contain significant nebulosity and requires that new sources detected in these regions be relatively sharp and PSF-like; sources significantly broadened by neighboring sources will not be modeled.  This process occasionally incorrectly identifies features in the wings of bright stars as nebulous, leading to fewer detected sources in these regions.  Only roughly 0.7\%  of the footprint is identified as being affected by nebulosity, however, so this is a small effect.  Likewise, some areas containing significant nebulosity are not marked and may contain many spurious sources.

\subsection{Uncertainty estimates}
\label{subsec:uncertaintylimitations}

The unWISE Catalog analysis adopts the unWISE coadd uncertainty images with little alteration, except near the ecliptic poles (\textsection\ref{subsec:eclipticpoles}).  Analysis of the coadd residual images after subtraction of the best fit models suggests that the coadd uncertainty images overestimate the actual uncertainty by 15\% in W1 and 20\% in W2.  On the other hand, the residual images show significant correlated uncertainties, which are neglected in the analysis.  If the correlated errors in the residuals could be eliminated or modeled out, and the uncertainty images modified to better describe the true dispersion in the values, WISE catalogs could reliably detect objects $\sim 0.2$ mag fainter than at present.

The unWISE Catalog reports formal statistical uncertainties that would be obtained for isolated point sources, without accounting for the covariance between blended sources.  The covariance between nearby sources will dominate the uncertainty of most sources in the bulge, for example, as well as faint sources near bright sources at all Galactic latitudes.  The importance of blending can be assessed via the catalog column \verb|fracflux|; values near 1 indicate that the source is isolated, while values near 0 indicate that most of the flux in the vicinity of this source comes from another object (Table~\ref{tab:detectioncatalog}).

The catalog also neglects systematic uncertainties in the measurement, which dominate the uncertainties of bright sources, even when isolated.  It is challenging to assess the size of the systematic uncertainty.  Were we to analyze the time-resolved unWISE coadd images, we could empirically measure the repeatability of the photometry for non-variable stars, but that is beyond the scope of this work.  Nevertheless, we can establish a floor on the systematic uncertainty by considering sources in the overlap regions between adjacent unWISE coadds, for which we have multiple unWISE catalog entries.  For sources brighter than 14th mag, the rms difference in the magnitudes of these sources is 3 mmag, possibly stemming from imperfect subpixel PSF interpolation, which dominates the residuals of bright stars (for example, Figure~\ref{fig:crowdsourceexample}).  Because the different unWISE coadds are drawn from identical underlying WISE individual-epoch images, this approach will strictly underestimate the true systematic error floor.  Additionally, this approach is not sensitive to systematic trends with magnitude (for example, \textsection\ref{subsec:nonlinearity}) or color. 

\section{Data Release}
\label{sec:release}

The unWISE coadds comprise 18240 tiles in two bands.  The corresponding 36480 catalog FITS files \citep{Pence:2010} composing the unWISE Catalog are available at the catalog web site, \url{http://catalog.unwise.me}.  The Catalog consists of 2.03 billion detections of objects in the ``primary'' regions of their unWISE coadds that have at least $5\sigma$ significance in W1 or W2.  Some basic numbers describing the catalog are given in Table~\ref{tab:numbers}.

\begin{deluxetable*}{crrrr}
\tablecaption{Number of Sources in the unWISE Catalog}
\tablehead{
\colhead{type} & \colhead{W1 \& W2} & \colhead{W1 only} & \colhead{W2 only} & \colhead{total}}
\startdata
\label{tab:numbers}
all                   &  1,063,569,639 &   1,032,419,887 &     118,744,698 &   2,214,734,224 \\
primary               &    979,399,857 &     949,169,093 &     108,858,251 &   2,037,427,201 \\
$>5\sigma$            &  1,062,021,630 &   1,027,533,192 &     116,683,978 &   2,206,238,800 \\
primary \& $>5\sigma$ &    978,015,749 &     944,811,706 &     106,974,415 &   2,029,801,870 \\
\enddata
\tablecomments{Number of objects in the unWISE Catalog satisfying various criteria.  ``primary'' sources discard duplicate sources in sky regions included in multiple coadds.  $>5\sigma$ sources are detected with at least $5\sigma$ significance.}
\end{deluxetable*}

The content of the FITS files is essentially identical to the files of the DECam Plane Survey \citep{Schlafly:2018}, with the addition of a few metadata columns and \texttt{spread\_model} (\textsection\ref{subsec:spreadmodel}).  The catalog columns are listed in Table~\ref{tab:detectioncatalog}.

\begin{deluxetable}{ll}
\tablewidth{\columnwidth}
\tablecaption{Catalog Columns}
\tablehead{
\colhead{Name} & \colhead{Description}}
\startdata
\label{tab:detectioncatalog}
\texttt{ra} & right ascension (deg) \\
\texttt{dec} & declination (deg) \\
\texttt{x} & $x$ coordinate (pix) \\
\texttt{y} & $y$ coordinate (pix) \\
\texttt{flux} & Vega flux (nMgy) \\
\texttt{dx} & $x$ uncertainty (pix) \\
\texttt{dy} & $y$ uncertainty (pix) \\
\texttt{dflux} & formal \texttt{flux} uncertainty (nMgy) \\
\texttt{fluxlbs} & local-background-subtracted flux (nMgy) \\
\texttt{dfluxlbs} & formal \texttt{fluxlbs} uncertainty (nMgy) \\
\hline
\texttt{qf} & PSF-weighted fraction of good pixels \\
\texttt{rchi2} & PSF-weighted average $\chi^2$ \\
\texttt{fracflux} & PSF-weighted fraction of flux from this source \\
\texttt{spread\_model} & SExtractor-like source size parameter \\
\texttt{dspread\_model} & uncertainty in \texttt{spread\_model} \\
\texttt{fwhm} & FWHM of PSF at source location (pix) \\
\texttt{sky} & residual sky at source location (nMgy) \\
\hline
\texttt{nm} & number of images in coadd at source \\
\texttt{primary} & source located in primary region of coadd \\
\texttt{flags\_unwise} & unWISE flags at source location \\
\texttt{flags\_info} & additional flags at source location \\
\hline
\texttt{coadd\_id} & unWISE/AllWISE \texttt{coadd\_id} of source \\
\texttt{band} & 1 for W1, 2 for W2 \\
\texttt{unwise\_detid} & detection ID, unique in catalog \\
\enddata
\tablecomments{
Columns in the unWISE catalogs.  A more complete description is available at the survey web site.
}
\end{deluxetable}

Fluxes and corresponding uncertainties are given in linear flux units, specifically, in Vega nanomaggies (nMgy) \citep{Finkbeiner:2004}.  The corresponding Vega magnitudes are given by $m_\mathrm{Vega} = 22.5-2.5\log_{10} \texttt{flux}$.  The following equations give the corresponding AB magnitudes:
\begin{align*}
m_\mathrm{W1,\, AB} &= m_\mathrm{W1,\, Vega} + 2.699 \\
m_\mathrm{W2,\, AB} &= m_\mathrm{W2,\, Vega} + 3.339 \, .
\end{align*}
As noted in \textsection\ref{subsec:photometry}, the agreement between unWISE and AllWISE magnitudes can be improved by subtracting 4 mmag and 32 mmag from W1 and W2.

Additional files give the \texttt{crowdsource} model image and sky image for each unWISE tile.  The PSF flux inverse variance image, mask image, and PSF model are also available for each tile.

The column \texttt{fwhm} is intended to give a sense of the size of the model PSF for a particular detection.  Given a PSF model, it is computed as the FWHM a Gaussian PSF with equal $n_\mathrm{eff}$ would have.  Typical values of 7.2\arcsec\ in W1 and 7.8\arcsec\ in W2 are larger than the true WISE FWHMs because the WISE PSF has more flux in its wings than a Gaussian, increasing $n_\mathrm{eff}$.

The column \texttt{primary} marks whether a particular source is located in the ``primary'' region of its coadd.  The unWISE coadds overlap one another by roughly 60 pixels, so that sources residing on the edges of an unWISE coadd will be detected in multiple coadds.  By selecting only ``primary'' sources, duplicate sources can be eliminated. Determining whether a source is primary in a given tile is purely a geometric operation, and does not involve any cross-matching of detections on neighboring tiles. For each source in a given tile's catalog, we compute the source's minimum distance from any edge of that tile's footprint. Using the source's (RA, Dec) coordinates, we perform the same minimum edge distance computation for all neighboring tiles. The source is labeled primary in the tile under consideration if that tile's footprint provides a larger minimum distance to any edge than do all other neighboring tile footprints.

Finally, merged catalogs linking W1 detections and W2 detections into multiband objects are also available.  Each W1 source is matched to the nearest W2 source within 2.4\arcsec; W2 sources within 2.4\arcsec\ that are not the closest source to a W1 source are considered unmatched.  The merged catalogs include the same columns as in the individual catalogs (Table~\ref{tab:detectioncatalog}), but each column now contains a two element vector for the W1 and W2 quantities.  The canonical right ascension and declination are taken to be the W1 quantities, when available, and otherwise the W2 quantities.  Likewise an object is considered ``primary'' when its W1 detection is considered ``primary.''  Finally, a unique \texttt{unwise\_objid} is assigned to each entry in the merged catalog.  Unmatched detections contain zeros in all columns corresponding to the missing band; these are easily identified, for example, by the empty \texttt{unwise\_detid}.

In additional to the catalogs themselves, the unWISE Catalog release contains a few items intended to facilitate the catalog's use.  These are
\begin{itemize}
    \item model images,
    \item model sky images,
    \item PSF depth images,
    \item mask images, and
    \item PSF images.
\end{itemize}
A sense for the plausibility of the modeling of a particular object in the unWISE Catalog can be obtained by comparing the unWISE coadds with the model images.  The unWISE model sky images and images of the PSF are potentially useful for users seeking to do their own photometry on the unWISE coadds; estimating the sky and PSF in crowded fields can be challenging.  Users wondering about spatial variations in the depth of the survey may find the PSF depth images valuable.

We also provide mask images that replicate some of the elements in the unWISE coadd mask images, but also add a few elements specific to the unWISE Catalog processing.  The values of the mask image are given in Table~\ref{tab:flags}.  In particular, the mask images indicate sky regions in which candidate sources significantly overlapping other sources are not modeled, due to the presence of a nearby large galaxy (\textsection\ref{subsec:nebulosity}).  They also indicate which parts of the sky the convolutional neural network indicates to be affected by significant nebulosity (\textsection\ref{subsec:nebulosity}).

\begin{deluxetable}{lll}
\tablewidth{\columnwidth}
\tablecaption{unWISE Catalog Info Flags}
\tablehead{
\colhead{Name} & \colhead{Bit} & \colhead{Description}}
\startdata
\label{tab:flags}
\texttt{bright\_off\_edge} & $2^0$ & bright source off coadd edge \\
\texttt{resolved\_galaxy} & $2^1$ & in large galaxy in HyperLeda \\
\texttt{big\_object} & $2^2$ & in M31 or Magellanic Cloud \\
\texttt{bright\_star\_cen} & $2^3$ & may contain bright star center \\
\texttt{crowdsat} & $2^4$ & may be affected by saturation \\
\texttt{nebulosity} & $2^5$ & nebulosity may be present \\
\texttt{nodeblend} & $2^6$ & deblending discouraged here \\
\texttt{sharp} & $2^7$ & only ``sharp'' sources here \\
\enddata
\tablecomments{
Informational flags in the unWISE Catalogs.  A more complete description is available at the survey web site.
}
\end{deluxetable}

\subsection{Source Designations}
We prescribe that unWISE Catalog source designations contain the prefix ``WISEU''. For example, WISEU J112234.53+122954.3 refers to the object with \verb|UNWISE_OBJID| = 1699p121o0017067.

\section{Conclusion}
\label{sec:conclusion}

The continuing NEOWISE-Reactivation mission has provided more than four years of imaging beyond the initial year of WISE data that was available for AllWISE processing.  The unWISE project has combined the $\sim25$ million single-frame WISE images into coadds reaching $2\times$ deeper than AllWISE.  The unWISE Catalog is the result of the analysis of these coadds.  It contains $\sim 2$ billion sources, roughly $3\times$ as many as cataloged in AllWISE.  Because of the broad WISE PSF, only a small fraction of extragalactic sources are resolved in WISE, making the analysis ideally suited to crowded-field pipelines that aggressively model images as sums of many overlapping point sources.  Application of the \texttt{crowdsource} crowded-field image analysis to the unWISE coadds provides accurate measurements for the unWISE Catalog, both in extragalactic fields where the improved depth of unWISE is critical, and also in the Galactic bulge where the analysis is limited by crowding.

Comparison between bright sources in the unWISE and AllWISE catalogs shows good agreement between the two catalogs.  For faint sources, comparison with deeper Spitzer imaging confirms that the catalog reaches $2\times$ deeper than AllWISE.  The unWISE Catalog reaches stars in the Milky Way at greater distances, detects hundreds of millions of new galaxies over $0 < z < 2$, and finds half of high-redshift quasars undetected in AllWISE.

We have outlined possible ways in which future versions of the unWISE Catalog may be able to augment or improve upon the present data products. Incorporating future NEOWISE data releases would enable the unWISE Catalog to push yet deeper. In combination with the complementary WISE-based proper motions that will be supplied by CatWISE (PI: Eisenhardt), the unWISE Catalog realizes much of the NEOWISE data set's tremendous potential for Galactic and extragalactic astrophysics.

unWISE coadd images, the derived unWISE catalog, the corresponding model PSF, sky images, and depth maps are publicly available at the unWISE web site, \url{http://unwise.me}, and the catalog web site, \url{http://catalog.unwise.me}.  

\vspace{5mm}

It is a pleasure to thank Roc Cutri for detailed feedback and John Moustakas for help with the HyperLeda catalog and the Legacy Survey Large Galaxy Atlas.  David Schlegel provided invaluable guidance and motivation, and Dustin Lang built much of the framework on which this catalog stands.  We thank the anonymous referee for valuable comments that improved the manuscript.

This work has been supported in part by NASA ADAP grant NNH17AE75I.  ES and AMM acknowledge support for this work provided by NASA through Hubble Fellowship grants HST-HF2-51367.001-A and HST-HF2-51415.001-A awarded by the Space Telescope Science Institute, which is operated by the Association of Universities for Research in Astronomy, Inc., for NASA, under contract NAS 5-26555.  ES and AMM acknowledge additional support by the Director, Office of Science, Office of High Energy Physics of the U.S. Department of Energy under Contract No. DE-AC02- 05CH11231, and by the National Energy Research Scientific Computing Center, a DOE Office of Science User Facility under the same contract 

We acknowledge the usage of the HyperLeda database (\url{http://leda.univ-lyon1.fr}).

This publication makes use of data products from the Wide-field Infrared Survey Explorer, which is a joint project of the University of California, Los Angeles, and the Jet Propulsion Laboratory/California Institute of Technology, and NEOWISE, which is a project of the Jet Propulsion Laboratory/California Institute of Technology. WISE and NEOWISE are funded by the National Aeronautics and Space Administration.

The unWISE Catalog analysis was run on the Odyssey cluster supported by the FAS Division of Science, Research Computing Group at Harvard University, and on the National Energy Research Scientific Computing Center, a DOE Office of Science User Facility supported by the Office of Science of the U.S. Department of Energy under Contract No. DE-AC02-05CH11231.  The unWISE nebulosity CNN was trained on the XStream computational resource, supported by the National Science Foundation Major Research Instrumentation program (ACI-1429830).

\appendix
\section{Nebulosity Neural Network Structure}
\label{app:neural-network-structure}

Table~\ref{tab:network-architecture} summarizes the architecture of the convolutional neural network that detects nebulosity in images. Because the input images are a quarter the size of those used in \citet{Schlafly:2018}, this network is slightly shallower.

We trained the network on three classes: \texttt{normal}, \texttt{nebulosity} and \texttt{nebulosity\_light}. We labeled 10,389 images by hand, using 80\% for training and the remaining 20\% for validation. After training, our network was able to separate cleanly between images labeled \texttt{normal} and \texttt{nebulosity}, with less than 0.1\% of images labeled \texttt{normal} being labeled \texttt{nebulosity} and vice versa. As the distinction between \texttt{nebulosity} and \texttt{nebulosity\_light} is more subtle (and likely less consistent in our hand classifications), our network misclassified 12\% of the images we considered \texttt{nebulosity} as \texttt{nebulosity\_light}.

\begin{deluxetable}{c|c|c}
    \tablecaption{Network architecture}
    \tablehead{\colhead{layer} & \colhead{output shape} & \colhead{details}
    \label{tab:network-architecture}
    }
    \startdata
        conv2d\_1 &
        $256 \times 256 , \ 12$ &
        $\begin{pmatrix}
        3 \times 3 , \, \mathrm{same \ padded} \\
        3 \times 3 , \, \mathrm{same \ padded} \\
        1 \times 1 \hphantom{, \, \mathrm{same \ padded}}
        \end{pmatrix}$ \\
        maxpool2d\_1 &
        $128 \times 128 , \ 12$ &
        $2 \times 2$ \\ \hline
        conv2d\_2 &
        $128 \times 128 , \ 16$ &
        $\begin{pmatrix}
        3 \times 3 , \, \mathrm{same \ padded} \\
        3 \times 3 , \, \mathrm{same \ padded} \\
        1 \times 1 \hphantom{, \, \mathrm{same \ padded}}
        \end{pmatrix}$ \\
        maxpool2d\_2 &
        $64 \times 64 , \ 16$ &
        $2 \times 2$ \\ \hline
        conv2d\_3 &
        $64 \times 64 , \ 24$ &
        $\begin{pmatrix}
        3 \times 3 , \, \mathrm{same \ padded} \\
        3 \times 3 , \, \mathrm{same \ padded} \\
        1 \times 1 \hphantom{, \, \mathrm{same \ padded}}
        \end{pmatrix}$ \\
        maxpool2d\_3 &
        $32 \times 32 , \, 24$ &
        $2 \times 2$ \\ \hline
        conv2d\_4 &
        $32 \times 32, \ 32$ &
        $\begin{pmatrix}
        3 \times 3 , \, \mathrm{same \ padded} \\
        3 \times 3 , \, \mathrm{same \ padded} \\
        1 \times 1 \hphantom{, \, \mathrm{same \ padded}}
        \end{pmatrix}$ \\
        maxpool2d\_4 &
        $16 \times 16 , \ 32$ &
        $2 \times 2$ \\ \hline
        conv2d\_5 &
        $16 \times 16, \ 32$ &
        $\begin{pmatrix}
        3 \times 3 , \, \mathrm{same \ padded} \\
        3 \times 3 , \, \mathrm{same \ padded} \\
        1 \times 1 \hphantom{, \, \mathrm{same \ padded}}
        \end{pmatrix}$ \\
        maxpool2d\_5 &
        $8 \times 8 , \ 32$ &
        $2 \times 2$ \\ \hline
        global\_avg\_pool2d &
        32 &
        \\ \hline
        dense\_1 &
        9 &
        20\% dropout \\
        dense\_2 &
        3 &
        10\% dropout \\ \hline
        softmax &
        3 & \\
        \enddata
    \tablecomments{All convolutional and dense layers use ReLU activation, and have an L2 weight penalty of $10^{-4}$. The final output one-hot encodes the class, and the cross-entropy loss function is used.}
\end{deluxetable}

\vspace{5mm}
\facilities{WISE}
\software{astropy \citep{astropy2013a}}

\bibliography{unwisecat}

\begin{thebibliography}{}
\expandafter\ifx\csname natexlab\endcsname\relax\def\natexlab#1{#1}\fi
\providecommand{\url}[1]{\href{#1}{#1}}

\bibitem[{{Antilogus} {et~al.}(2014){Antilogus}, {Astier}, {Doherty},
  {Guyonnet}, \& {Regnault}}]{Antilogus:2014}
{Antilogus}, P., {Astier}, P., {Doherty}, P., {Guyonnet}, A., \& {Regnault}, N.
  2014, Journal of Instrumentation, 9, C03048

\bibitem[{{Astropy Collaboration} {et~al.}(2013){Astropy Collaboration},
  {Robitaille}, {Tollerud}, {Greenfield}, {Droettboom}, {Bray}, {Aldcroft},
  {Davis}, {Ginsburg}, {Price-Whelan}, {Kerzendorf}, {Conley}, {Crighton},
  {Barbary}, {Muna}, {Ferguson}, {Grollier}, {Parikh}, {Nair}, {Unther},
  {Deil}, {Woillez}, {Conseil}, {Kramer}, {Turner}, {Singer}, {Fox}, {Weaver},
  {Zabalza}, {Edwards}, {Azalee Bostroem}, {Burke}, {Casey}, {Crawford},
  {Dencheva}, {Ely}, {Jenness}, {Labrie}, {Lim}, {Pierfederici}, {Pontzen},
  {Ptak}, {Refsdal}, {Servillat}, \& {Streicher}}]{astropy2013a}
{Astropy Collaboration}, {Robitaille}, T.~P., {Tollerud}, E.~J., {et~al.} 2013,
  \aap, 558, A33

\bibitem[{{Ba{\~n}ados} {et~al.}(2018){Ba{\~n}ados}, {Venemans},
  {Mazzucchelli}, {Farina}, {Walter}, {Wang}, {Decarli}, {Stern}, {Fan},
  {Davies}, {Hennawi}, {Simcoe}, {Turner}, {Rix}, {Yang}, {Kelson}, {Rudie}, \&
  {Winters}}]{banados2018}
{Ba{\~n}ados}, E., {Venemans}, B.~P., {Mazzucchelli}, C., {et~al.} 2018, \nat,
  553, 473

\bibitem[{{Benjamin} {et~al.}(2003){Benjamin}, {Churchwell}, {Babler}, {Bania},
  {Clemens}, {Cohen}, {Dickey}, {Indebetouw}, {Jackson}, {Kobulnicky},
  {Lazarian}, {Marston}, {Mathis}, {Meade}, {Seager}, {Stolovy}, {Watson},
  {Whitney}, {Wolff}, \& {Wolfire}}]{Benjamin:2003}
{Benjamin}, R.~A., {Churchwell}, E., {Babler}, B.~L., {et~al.} 2003, \pasp,
  115, 953

\bibitem[{{Bertin} \& {Arnouts}(1996)}]{Bertin:1996}
{Bertin}, E., \& {Arnouts}, S. 1996, \aaps, 117, 393

\bibitem[{{Churchwell} {et~al.}(2009){Churchwell}, {Babler}, {Meade},
  {Whitney}, {Benjamin}, {Indebetouw}, {Cyganowski}, {Robitaille}, {Povich},
  {Watson}, \& {Bracker}}]{Churchwell:2009}
{Churchwell}, E., {Babler}, B.~L., {Meade}, M.~R., {et~al.} 2009, \pasp, 121,
  213

\bibitem[{{Cutri} {et~al.}(2013){Cutri}, {Wright}, {Conrow}, {Fowler},
  {Eisenhardt}, {Grillmair}, {Kirkpatrick}, {Masci}, {McCallon}, {Wheelock},
  {Fajardo-Acosta}, {Yan}, {Benford}, {Harbut}, {Jarrett}, {Lake}, {Leisawitz},
  {Ressler}, {Stanford}, {Tsai}, {Liu}, {Helou}, {Mainzer}, {Gettings},
  {Gonzalez}, {Hoffman}, {Marsh}, {Padgett}, {Skrutskie}, {Beck}, {Papin}, \&
  {Wittman}}]{Cutri:2013}
{Cutri}, R.~M., {Wright}, E.~L., {Conrow}, T., {et~al.} 2013, {Explanatory
  Supplement to the AllWISE Data Release Products}, Tech. rep.

\bibitem[{{Cutri} {et~al.}(2015){Cutri}, {Mainzer}, {Conrow}, {Masci}, {Bauer},
  {Dailey}, {Kirkpatrick}, {Fajardo-Acosta}, {Gelino}, {Grillmair}, {Wheelock},
  {Yan}, {Harbut}, {Beck}, {Wittman}, {Wright}, {Masiero}, {Grav}, {Sonnett},
  {Nugent}, {Kramer}, {Stevenson}, {Eisenhardt}, {Fabinsky}, {Tholen}, {Papin},
  {Fowler}, \& {McCallon}}]{neowise_supplement}
{Cutri}, R.~M., {Mainzer}, A., {Conrow}, T., {et~al.} 2015, {Explanatory
  Supplement to the NEOWISE Data Release Products}, Tech. rep.

\bibitem[{Cypriano {et~al.}(2010)Cypriano, Amara, Voigt, Bridle, Abdalla,
  Réfrégier, Seiffert, \& Rhodes}]{Cypriano:2010}
Cypriano, E.~S., Amara, A., Voigt, L.~M., {et~al.} 2010, Monthly Notices of the
  Royal Astronomical Society, 405, 494.
\newblock \url{http://dx.doi.org/10.1111/j.1365-2966.2010.16461.x}

\bibitem[{{Desai} {et~al.}(2012){Desai}, {Armstrong}, {Mohr}, {Semler}, {Liu},
  {Bertin}, {Allam}, {Barkhouse}, {Bazin}, {Buckley-Geer}, {Cooper}, {Hansen},
  {High}, {Lin}, {Lin}, {Ngeow}, {Rest}, {Song}, {Tucker}, \&
  {Zenteno}}]{Desai:2012}
{Desai}, S., {Armstrong}, R., {Mohr}, J.~J., {et~al.} 2012, \apj, 757, 83

\bibitem[{{Dey} {et~al.}(2018){Dey}, {Schlegel}, {Lang}, {Blum}, {Burleigh},
  {Fan}, {Findlay}, {Finkbeiner}, {Herrera}, {Juneau}, {Landriau}, {Levi},
  {McGreer}, {Meisner}, {Myers}, {Moustakas}, {Nugent}, {Patej}, {Schlafly},
  {Walker}, {Valdes}, {Weaver}, {Yeche}, {Zou}, {Zhou}, {Abareshi}, {Abbott},
  {Abolfathi}, {Aguilera}, {Allen}, {Alvarez}, {Annis}, {Aubert}, {Bell},
  {BenZvi}, {Bielby}, {Bolton}, {Briceno}, {Buckley-Geer}, {Butler},
  {Calamida}, {Carlberg}, {Carter}, {Casas}, {Castander}, {Choi}, {Comparat},
  {Cukanovaite}, {Delubac}, {DeVries}, {Dey}, {Dhungana}, {Dickinson}, {Ding},
  {Donaldson}, {Duan}, {Duckworth}, {Eftekharzadeh}, {Eisenstein}, {Etourneau},
  {Fagrelius}, {Farihi}, {Fitzpatrick}, {Font-Ribera}, {Fulmer}, {Gansicke},
  {Gaztanaga}, {George}, {Gerdes}, {Gontcho}, {Green}, {Guy}, {Harmer},
  {Hernandez}, {Honscheid}, {Lijuan}, {Huang}, {James}, {Jannuzi}, {Jiang},
  {Joyce}, {Karcher}, {Karkar}, {Kehoe}, {Kneib}, {Kueter-Young}, {Lan},
  {Lauer}, {Le Guillou}, {Le Van Suu}, {Lee}, {Lesser}, {Li}, {Mann},
  {Marshall}, {Mart{\'{\i}}nez-V{\'a}zquez}, {Martini}, {du Mas des Bourboux},
  {McManus}, {Menard}, {Metcalfe}, {Mu{\~n}oz-Guti{\'e}rrez}, {Najita},
  {Napier}, {Narayan}, {Newman}, {Nie}, {Nord}, {Norman}, {Olsen}, {Paat},
  {Palanque-Delabrouille}, {Peng}, {Poppett}, {Poremba}, {Prakash},
  {Rabinowitz}, {Raichoor}, {Rezaie}, {Robertson}, {Roe}, {Ross}, {Ross},
  {Rudnick}, {Safonova}, {Saha}, {Sanchez}, {Schweiker}, {Scott}, {Seo},
  {Shan}, {Silva}, {Soto}, {Sprayberry}, {Staten}, {Stillman}, {Stupak},
  {Summers}, {Sien Tie}, {Tirado}, {Vargas-Magana}, {Vivas}, {Wechsler},
  {Williams}, {Yang}, {Yang}, {Yapici}, {Zaritsky}, {Zenteno}, {Zhang},
  {Zhang}, {Zhou}, \& {Zhou}}]{dey2018}
{Dey}, A., {Schlegel}, D.~J., {Lang}, D., {et~al.} 2018, ArXiv e-prints,
  arXiv:1804.08657

\bibitem[{{Downing} {et~al.}(2006){Downing}, {Baade}, {Sinclaire}, {Deiries},
  \& {Christen}}]{Downing:2006}
{Downing}, M., {Baade}, D., {Sinclaire}, P., {Deiries}, S., \& {Christen}, F.
  2006, in \procspie, Vol. 6276, Society of Photo-Optical Instrumentation
  Engineers (SPIE) Conference Series, 627609

\bibitem[{{Finkbeiner} {et~al.}(2004){Finkbeiner}, {Padmanabhan}, {Schlegel},
  {Carr}, {Gunn}, {Rockosi}, {Sekiguchi}, {Lupton}, {Knapp}, {Ivezi{\'c}},
  {Blanton}, {Hogg}, {Adelman-McCarthy}, {Annis}, {Hayes}, {Kinney}, {Long},
  {Seljak}, {Strauss}, {Yanny}, {Ag{\"u}eros}, {Allam}, {Anderson}, {Bahcall},
  {Baldry}, {Bernardi}, {Boroski}, {Briggs}, {Brinkmann}, {Brunner},
  {Budav{\'a}ri}, {Castander}, {Covey}, {Csabai}, {Doi}, {Dong}, {Eisenstein},
  {Fan}, {Friedman}, {Fukugita}, {Gillespie}, {Grebel}, {Gurbani}, {de Haas},
  {Harris}, {Hendry}, {Hennessy}, {Jester}, {Johnston}, {Jorgensen},
  {Juri{\'c}}, {Kent}, {Kniazev}, {Krzesi{\'n}ski}, {Leger}, {Lin}, {Loveday},
  {Mannery}, {Mart{\'{\i}}nez-Delgado}, {McGehee}, {Meiksin}, {Munn},
  {Neilsen}, {Newman}, {Nitta}, {Pauls}, {Quinn}, {Rafikov}, {Richards},
  {Richmond}, {Schneider}, {Schroeder}, {Shimasaku}, {Siegmund}, {Smith},
  {Snedden}, {Stebbins}, {Szalay}, {Szokoly}, {Tegmark}, {Tucker}, {Uomoto},
  {Vanden Berk}, {Weinberg}, {West}, {Yasuda}, {Yocum}, {York}, \&
  {Zehavi}}]{Finkbeiner:2004}
{Finkbeiner}, D.~P., {Padmanabhan}, N., {Schlegel}, D.~J., {et~al.} 2004, \aj,
  128, 2577

\bibitem[{{Gaia Collaboration} {et~al.}(2016){Gaia Collaboration}, {Prusti},
  {de Bruijne}, {Brown}, {Vallenari}, {Babusiaux}, {Bailer-Jones}, {Bastian},
  {Biermann}, {Evans}, \& et~al.}]{Gaia:2016}
{Gaia Collaboration}, {Prusti}, T., {de Bruijne}, J.~H.~J., {et~al.} 2016,
  \aap, 595, A1

\bibitem[{{Kirkpatrick} {et~al.}(2011){Kirkpatrick}, {Cushing}, {Gelino},
  {Griffith}, {Skrutskie}, {Marsh}, {Wright}, {Mainzer}, {Eisenhardt},
  {McLean}, {Thompson}, {Bauer}, {Benford}, {Bridge}, {Lake}, {Petty},
  {Stanford}, {Tsai}, {Bailey}, {Beichman}, {Bloom}, {Bochanski}, {Burgasser},
  {Capak}, {Cruz}, {Hinz}, {Kartaltepe}, {Knox}, {Manohar}, {Masters},
  {Morales-Calder{\'o}n}, {Prato}, {Rodigas}, {Salvato}, {Schurr}, {Scoville},
  {Simcoe}, {Stapelfeldt}, {Stern}, {Stock}, \& {Vacca}}]{kirkpatrick2011}
{Kirkpatrick}, J.~D., {Cushing}, M.~C., {Gelino}, C.~R., {et~al.} 2011, \apjs,
  197, 19

\bibitem[{{Laigle} {et~al.}(2016){Laigle}, {McCracken}, {Ilbert}, {Hsieh},
  {Davidzon}, {Capak}, {Hasinger}, {Silverman}, {Pichon}, {Coupon}, {Aussel},
  {Le Borgne}, {Caputi}, {Cassata}, {Chang}, {Civano}, {Dunlop}, {Fynbo},
  {Kartaltepe}, {Koekemoer}, {Le F{\`e}vre}, {Le Floc'h}, {Leauthaud}, {Lilly},
  {Lin}, {Marchesi}, {Milvang-Jensen}, {Salvato}, {Sanders}, {Scoville},
  {Smolcic}, {Stockmann}, {Taniguchi}, {Tasca}, {Toft}, {Vaccari}, \&
  {Zabl}}]{Laigle:2016}
{Laigle}, C., {McCracken}, H.~J., {Ilbert}, O., {et~al.} 2016, \apjs, 224, 24

\bibitem[{{Lang}(2014)}]{lang_unwise_coadds}
{Lang}, D. 2014, \aj, 147, 108

\bibitem[{{Lang} {et~al.}(2014){Lang}, {Hogg}, \&
  {Schlegel}}]{unwise_sdss_forcedphot}
{Lang}, D., {Hogg}, D.~W., \& {Schlegel}, D.~J. 2014, ArXiv e-prints,
  arXiv:1410.7397

\bibitem[{{Low} {et~al.}(2007){Low}, {Rieke}, \& {Gehrz}}]{low2007}
{Low}, F.~J., {Rieke}, G.~H., \& {Gehrz}, R.~D. 2007, \araa, 45, 43

\bibitem[{{Mainzer} {et~al.}(2011){Mainzer}, {Bauer}, {Grav}, {Masiero},
  {Cutri}, {Dailey}, {Eisenhardt}, {McMillan}, {Wright}, {Walker}, {Jedicke},
  {Spahr}, {Tholen}, {Alles}, {Beck}, {Brandenburg}, {Conrow}, {Evans},
  {Fowler}, {Jarrett}, {Marsh}, {Masci}, {McCallon}, {Wheelock}, {Wittman},
  {Wyatt}, {DeBaun}, {Elliott}, {Elsbury}, {Gautier}, {Gomillion}, {Leisawitz},
  {Maleszewski}, {Micheli}, \& {Wilkins}}]{neowise}
{Mainzer}, A., {Bauer}, J., {Grav}, T., {et~al.} 2011, \apj, 731, 53

\bibitem[{{Mainzer} {et~al.}(2014){Mainzer}, {Bauer}, {Cutri}, {Grav},
  {Masiero}, {Beck}, {Clarkson}, {Conrow}, {Dailey}, {Eisenhardt}, {Fabinsky},
  {Fajardo-Acosta}, {Fowler}, {Gelino}, {Grillmair}, {Heinrichsen}, {Kendall},
  {Kirkpatrick}, {Liu}, {Masci}, {McCallon}, {Nugent}, {Papin}, {Rice},
  {Royer}, {Ryan}, {Sevilla}, {Sonnett}, {Stevenson}, {Thompson}, {Wheelock},
  {Wiemer}, {Wittman}, {Wright}, \& {Yan}}]{neowiser}
{Mainzer}, A., {Bauer}, J., {Cutri}, R.~M., {et~al.} 2014, \apj, 792, 30

\bibitem[{{Makarov} {et~al.}(2014){Makarov}, {Prugniel}, {Terekhova},
  {Courtois}, \& {Vauglin}}]{Makarov:2014}
{Makarov}, D., {Prugniel}, P., {Terekhova}, N., {Courtois}, H., \& {Vauglin},
  I. 2014, \aap, 570, A13

\bibitem[{{Meisner} \& {Finkbeiner}(2014)}]{wise_dust_map}
{Meisner}, A.~M., \& {Finkbeiner}, D.~P. 2014, \apj, 781, 5

\bibitem[{{Meisner} {et~al.}(2017{\natexlab{a}}){Meisner}, {Lang}, \&
  {Schlegel}}]{fulldepth_neo1}
{Meisner}, A.~M., {Lang}, D., \& {Schlegel}, D.~J. 2017{\natexlab{a}}, \aj,
  153, 38

\bibitem[{{Meisner} {et~al.}(2017{\natexlab{b}}){Meisner}, {Lang}, \&
  {Schlegel}}]{fulldepth_neo2}
---. 2017{\natexlab{b}}, \aj, 154, 161

\bibitem[{{Meisner} {et~al.}(2018{\natexlab{a}}){Meisner}, {Lang}, \&
  {Schlegel}}]{fulldepth_neo3}
---. 2018{\natexlab{a}}, Research Notes of the American Astronomical Society,
  2, 1

\bibitem[{{Meisner} {et~al.}(2018{\natexlab{b}}){Meisner}, {Lang}, \&
  {Schlegel}}]{tr_neo2}
---. 2018{\natexlab{b}}, \aj, 156, 69

\bibitem[{{Meisner} {et~al.}(2018{\natexlab{c}}){Meisner}, {Lang}, \&
  {Schlegel}}]{tr_neo3}
{Meisner}, A.~M., {Lang}, D.~A., \& {Schlegel}, D.~J. 2018{\natexlab{c}},
  Research Notes of the American Astronomical Society, 2, 202

\bibitem[{{Pence} {et~al.}(2010){Pence}, {Chiappetti}, {Page}, {Shaw}, \&
  {Stobie}}]{Pence:2010}
{Pence}, W.~D., {Chiappetti}, L., {Page}, C.~G., {Shaw}, R.~A., \& {Stobie}, E.
  2010, \aap, 524, A42

\bibitem[{{Sanders} {et~al.}(2007{\natexlab{a}}){Sanders}, {Salvato}, {Aussel},
  {Ilbert}, {Scoville}, {Surace}, {Frayer}, {Sheth}, {Helou}, {Brooke},
  {Bhattacharya}, {Yan}, {Kartaltepe}, {Barnes}, {Blain}, {Calzetti}, {Capak},
  {Carilli}, {Carollo}, {Comastri}, {Daddi}, {Ellis}, {Elvis}, {Fall},
  {Franceschini}, {Giavalisco}, {Hasinger}, {Impey}, {Koekemoer}, {Le
  F{\`e}vre}, {Lilly}, {Liu}, {McCracken}, {Mobasher}, {Renzini}, {Rich},
  {Schinnerer}, {Shopbell}, {Taniguchi}, {Thompson}, {Urry}, \&
  {Williams}}]{Sanders:2007}
{Sanders}, D.~B., {Salvato}, M., {Aussel}, H., {et~al.} 2007{\natexlab{a}},
  \apjs, 172, 86

\bibitem[{{Sanders} {et~al.}(2007{\natexlab{b}}){Sanders}, {Salvato}, {Aussel},
  {Ilbert}, {Scoville}, {Surace}, {Frayer}, {Sheth}, {Helou}, {Brooke},
  {Bhattacharya}, {Yan}, {Kartaltepe}, {Barnes}, {Blain}, {Calzetti}, {Capak},
  {Carilli}, {Carollo}, {Comastri}, {Daddi}, {Ellis}, {Elvis}, {Fall},
  {Franceschini}, {Giavalisco}, {Hasinger}, {Impey}, {Koekemoer}, {Le
  F{\`e}vre}, {Lilly}, {Liu}, {McCracken}, {Mobasher}, {Renzini}, {Rich},
  {Schinnerer}, {Shopbell}, {Taniguchi}, {Thompson}, {Urry}, \&
  {Williams}}]{scosmos}
---. 2007{\natexlab{b}}, \apjs, 172, 86

\bibitem[{{Schlafly} {et~al.}(2017){Schlafly}, {Peek}, {Finkbeiner}, \&
  {Green}}]{Schlafly:2017}
{Schlafly}, E.~F., {Peek}, J.~E.~G., {Finkbeiner}, D.~P., \& {Green}, G.~M.
  2017, \apj, 838, 36

\bibitem[{{Schlafly} {et~al.}(2016){Schlafly}, {Meisner}, {Stutz},
  {Kainulainen}, {Peek}, {Tchernyshyov}, {Rix}, {Finkbeiner}, {Covey}, {Green},
  {Bell}, {Burgett}, {Chambers}, {Draper}, {Flewelling}, {Hodapp}, {Kaiser},
  {Magnier}, {Martin}, {Metcalfe}, {Wainscoat}, \& {Waters}}]{Schlafly:2016}
{Schlafly}, E.~F., {Meisner}, A.~M., {Stutz}, A.~M., {et~al.} 2016, \apj, 821,
  78

\bibitem[{{Schlafly} {et~al.}(2018){Schlafly}, {Green}, {Lang}, {Daylan},
  {Finkbeiner}, {Lee}, {Meisner}, {Schlegel}, \& {Valdes}}]{Schlafly:2018}
{Schlafly}, E.~F., {Green}, G.~M., {Lang}, D., {et~al.} 2018, \apjs, 234, 39

\bibitem[{{Schlegel} {et~al.}(2015){Schlegel}, {Blum}, {Castander}, {Dey},
  {Finkbeiner}, {Foucaud}, {Honscheid}, {James}, {Lang}, {Levi}, {Moustakas},
  {Myers}, {Newman}, {Nord}, {Nugent}, {Patej}, {Reil}, {Rudnick}, {Rykoff},
  {Ford Schlafly}, {Stark}, {Valdes}, {Walker}, {Weaver}, \& {DECam Legacy
  Survey Collaboration}}]{schlegel2015}
{Schlegel}, D.~J., {Blum}, R.~D., {Castander}, F.~J., {et~al.} 2015, in
  American Astronomical Society Meeting Abstracts, Vol. 225, American
  Astronomical Society Meeting Abstracts \#225, 336.07

\bibitem[{{Wang} {et~al.}(2016){Wang}, {Wu}, {Fan}, {Yang}, {Yi}, {Bian},
  {McGreer}, {Yang}, {Ai}, {Dong}, {Zuo}, {Jiang}, {Green}, {Wang}, {Cai},
  {Wang}, \& {Yue}}]{Wang:2016}
{Wang}, F., {Wu}, X.-B., {Fan}, X., {et~al.} 2016, \apj, 819, 24

\bibitem[{{Wheelock} {et~al.}(1994){Wheelock}, {Gautier}, {Chillemi}, {Kester},
  {McCallon}, {Oken}, {White}, {Gregorich}, {Boulanger}, \&
  {Good}}]{wheelock1994}
{Wheelock}, S.~L., {Gautier}, T.~N., {Chillemi}, J., {et~al.} 1994, NASA
  STI/Recon Technical Report N, 95

\bibitem[{{Wright} {et~al.}(2010){Wright}, {Eisenhardt}, {Mainzer}, {Ressler},
  {Cutri}, {Jarrett}, {Kirkpatrick}, {Padgett}, {McMillan}, {Skrutskie},
  {Stanford}, {Cohen}, {Walker}, {Mather}, {Leisawitz}, {Gautier}, {McLean},
  {Benford}, {Lonsdale}, {Blain}, {Mendez}, {Irace}, {Duval}, {Liu}, {Royer},
  {Heinrichsen}, {Howard}, {Shannon}, {Kendall}, {Walsh}, {Larsen}, {Cardon},
  {Schick}, {Schwalm}, {Abid}, {Fabinsky}, {Naes}, \& {Tsai}}]{wright2010}
{Wright}, E.~L., {Eisenhardt}, P.~R.~M., {Mainzer}, A.~K., {et~al.} 2010, \aj,
  140, 1868

\bibitem[{{York} {et~al.}(2000){York}, {Adelman}, {Anderson}, {Anderson},
  {Annis}, {Bahcall}, {Bakken}, {Barkhouser}, {Bastian}, {Berman}, {Boroski},
  {Bracker}, {Briegel}, {Briggs}, {Brinkmann}, {Brunner}, {Burles}, {Carey},
  {Carr}, {Castander}, {Chen}, {Colestock}, {Connolly}, {Crocker}, {Csabai},
  {Czarapata}, {Davis}, {Doi}, {Dombeck}, {Eisenstein}, {Ellman}, {Elms},
  {Evans}, {Fan}, {Federwitz}, {Fiscelli}, {Friedman}, {Frieman}, {Fukugita},
  {Gillespie}, {Gunn}, {Gurbani}, {de Haas}, {Haldeman}, {Harris}, {Hayes},
  {Heckman}, {Hennessy}, {Hindsley}, {Holm}, {Holmgren}, {Huang}, {Hull},
  {Husby}, {Ichikawa}, {Ichikawa}, {Ivezi{\'c}}, {Kent}, {Kim}, {Kinney},
  {Klaene}, {Kleinman}, {Kleinman}, {Knapp}, {Korienek}, {Kron}, {Kunszt},
  {Lamb}, {Lee}, {Leger}, {Limmongkol}, {Lindenmeyer}, {Long}, {Loomis},
  {Loveday}, {Lucinio}, {Lupton}, {MacKinnon}, {Mannery}, {Mantsch}, {Margon},
  {McGehee}, {McKay}, {Meiksin}, {Merelli}, {Monet}, {Munn}, {Narayanan},
  {Nash}, {Neilsen}, {Neswold}, {Newberg}, {Nichol}, {Nicinski}, {Nonino},
  {Okada}, {Okamura}, {Ostriker}, {Owen}, {Pauls}, {Peoples}, {Peterson},
  {Petravick}, {Pier}, {Pope}, {Pordes}, {Prosapio}, {Rechenmacher}, {Quinn},
  {Richards}, {Richmond}, {Rivetta}, {Rockosi}, {Ruthmansdorfer}, {Sandford},
  {Schlegel}, {Schneider}, {Sekiguchi}, {Sergey}, {Shimasaku}, {Siegmund},
  {Smee}, {Smith}, {Snedden}, {Stone}, {Stoughton}, {Strauss}, {Stubbs},
  {SubbaRao}, {Szalay}, {Szapudi}, {Szokoly}, {Thakar}, {Tremonti}, {Tucker},
  {Uomoto}, {Vanden Berk}, {Vogeley}, {Waddell}, {Wang}, {Watanabe},
  {Weinberg}, {Yanny}, {Yasuda}, \& {SDSS Collaboration}}]{York:2000}
{York}, D.~G., {Adelman}, J., {Anderson}, Jr., J.~E., {et~al.} 2000, \aj, 120,
  1579

\end{thebibliography}
\end{document}